\documentclass{article}

\usepackage{graphicx}
\usepackage{amssymb}
\usepackage{amsmath}
\usepackage{amsthm}
\usepackage{mathrsfs}
\usepackage{url}
\usepackage{color}
\usepackage{longtable}
\usepackage{rotating}
\usepackage{authblk}
\usepackage{multirow}
\usepackage[authoryear]{natbib}

\definecolor{red}{rgb}{1,0,0}

\begin{document}

\title{Queue Imbalance as a One-Tick-Ahead Price Predictor in a Limit Order Book}

\author{Martin D. Gould\footnote{Corresponding author. Email: m.gould@imperial.ac.uk.} $^{\ddag}$ and Julius Bonart$^{\ddag}$}

\affil{$\ddag$ CFM--Imperial Institute of Quantitative Finance, Department of Mathematics, Imperial College, London SW7 2AZ}

\maketitle

\begin{abstract}We investigate whether the bid/ask queue imbalance in a limit order book (LOB) provides significant predictive power for the direction of the next mid-price movement. We consider this question both in the context of a simple binary classifier, which seeks to predict the direction of the next mid-price movement, and a probabilistic classifier, which seeks to predict the probability that the next mid-price movement will be upwards. To implement these classifiers, we fit logistic regressions between the queue imbalance and the direction of the subsequent mid-price movement for each of 10 liquid stocks on Nasdaq. In each case, we find a strongly statistically significant relationship between these variables. Compared to a simple null model, which assumes that the direction of mid-price changes is uncorrelated with the queue imbalance, we find that our logistic regression fits provide a considerable improvement in binary and probabilistic classification for large-tick stocks, and provide a moderate improvement in binary and probabilistic classification for small-tick stocks. We also perform local logistic regression fits on the same data, and find that this semi-parametric approach slightly outperform our logistic regression fits, at the expense of being more computationally intensive to implement.\end{abstract}

\textbf{Keywords: }Price prediction; queue imbalance; high-frequency trading; limit order books; market microstructure.

\section{Introduction}
 
In most modern financial markets, trade occurs via a continuous double-auction mechanism called a limit order book (LOB) \citep{Gould:2013limit}. In an LOB, traders interact by submitting orders that state their desires to buy or sell a specified quantity of an asset at a specified price. Active orders reside in a queue until they are either cancelled by their owner or executed against an order of opposite type. Thanks to electronic LOB trading platforms, traders from around the world can monitor the quantities that are available for purchase or sale at specified prices, and can thereby deduce a detailed, up-to-date picture of market state.

Since the widespread uptake of LOB trading, the question of whether information about LOB state can be used to formulate predictions of future price movements has remained a topic of primary interest to practitioners and researchers alike. Due to the potentially lucrative benefits of success, the topic has attracted the attention of countless professional and novice traders, who have sought to reap the financial rewards of discovering strategies that successfully forecast future prices. The question has similarly attracted the attention of many scholars from a wide range of disciplines, who have attempted to quantify several important empirical properties of price series \citep{Chakraborti:2011empirical,Cont:2001empirical}, to understand the origins and nature of price movements \citep{Bouchaud:2009digest,Farmer:2006market}, and to propose and evaluate models designed to explain the market dynamics from which price movements emerge \citep{Farmer:2005predictive,Gould:2013limit}.

A key difficulty in using information about LOB state to predict future price movements is the need to identify which state variables to use as inputs when making predictions. In recent years, some authors (see, e.g., \citet{Cartea:2015enhancing} and \citet{Yang:2015reduced}) have proposed that the \emph{queue imbalance}, which describes the difference between the volumes offered for purchase or sale at the best bid and ask quotes in a limit order book (LOB), could constitute a simple yet powerful quantity that is suitable for this purpose. Despite reasonably widespread discussion of this idea among practitioners, detailed, scientific analysis of the true predictive power of this measure has remained limited. Given the apparent prevalence of this approach among many real traders, developing a detailed understanding of both its statistical performance and the possible market dynamics underlying its success is an important and timely task.

In this paper, we study a recent, high-quality data set that describes the LOB activity for each of 10 liquid stocks on Nasdaq to assess whether the queue imbalance provides significant predictive power for the direction of the next mid-price movement. We consider this question both in the context of a simple binary classifier, which seeks to predict the direction of the next mid-price movement, and a probabilistic classifier, which seeks to predict the probability that the next mid-price movement will be upwards.

To implement these classifiers, we fit logistic regressions between the queue imbalance and the direction of the subsequent mid-price movement for each of the stocks in our sample. We ensure that our results are statistically and scientifically rigorous by implementing formal hypothesis tests and quantitative performance measures of out-of-sample forecasting. For each of the 10 stocks in our sample, we test and strongly reject the hypothesis that the fitted regression curve does not find a statistically significant relationship between queue imbalance and the direction of the subsequent mid-price movement.

We also introduce a simple null model, which assumes that the direction of mid-price changes is uncorrelated with the queue imbalance. Compared to the null model, we find that our logistic regressions improve out-of-sample performance of binary classification by about $50$--$60\%$ for large tick stocks and about $10$--$30\%$ for small-tick stocks. We find that our results for probabilistic predictions are slightly weaker, but still improve out-of-sample predictive performance by about $20$--$30\%$ for large-tick stocks and about $2$--$6\%$ for small-tick stocks.

Although our logistic regression fits indicate that queue imbalance provides significant predictive power for price prediction, the parametric nature of this approach obscures the detailed market dynamics that relate queue imbalance to subsequent mid-price movements. To help address this problem, we also complement our logistic regressions with a semi-parametric approach, by fitting local logistic regression curves to the same data. In contrast to the logistic regression fits, whose shapes are constrained by the parametric form of the logistic sigmoid function, our local logistic fits illustrate a more subtle relationship between queue imbalance and mid-price movements, and thereby help to illuminate how the market microstructure underpins our results. We find that our local logistic regressions slightly outperform our logistic regressions for both binary classification and probabilistic prediction, although we note that the strength of this improvement is relatively small given the considerable increase in computational power required to fit a local logistic regression. To conclude, we discuss how several important differences in LOB behaviour could help to explain the differences that we observe between small- and large-tick stocks, and discuss many possible avenues for future research.

There are many practical applications for the findings that we present. First, the ability to predict future price movements is a valuable tool for practitioners seeking to design trading strategies in financial markets. Many financial institutions invest vast sums of money to improve the predictive power of their forecasts by a tiny fraction of a percentage point, whereas the prediction methods that we present are extremely simple to implement and outperform our simple null model by a considerable margin. Second, thanks to their computational simplicity, the methods that we discuss can also be employed by electronic trading algorithms in real time, to help improve their performance. Third, the ability to formulate accurate forecasts of future price movements could be useful for traders who seek to implement optimal execution algorithms by deciding between submitting limit or market orders. Fourth, the empirical framework that we consider is also a useful laboratory in which to test models. Specifically, comparing a model's price predictions for a given queue imbalance to the corresponding result fitted directly from data could be useful as an indicator for how the model needs to be improved, or to rule it out altogether.

The paper proceeds as follows. In Section \ref{sec:lobs}, we provide a detailed description of price formation in an LOB. In Section \ref{sec:priceprediction}, we discuss several other publications that have addressed price prediction in an LOB, and we define the measures and terminology that we use for our own study. In Section \ref{sec:data}, we describe the data that forms the basis of our empirical calculations. In Section \ref{sec:methodology}, we discuss our statistical methodology. We present our main results in Section \ref{sec:results} and discuss our findings in Section \ref{sec:discussion}. Section \ref{sec:conclusions} concludes.

\section{Order Queues and Price Changes in a Limit Order Book}\label{sec:lobs}

\subsection{Limit Order Books}

More than half of the world's financial markets use electronic limit order books (LOBs) to facilitate trade \citep{Rosu:2009dynamic}. In contrast to quote-driven systems, in which prices are set by designated market makers, trade in an LOB occurs via a continuous double-auction mechanism whereby institutions submit orders. An \emph{order} $x=(p_x,\omega_x,t_x)$ submitted at time $t_x$ with price $p_x$ and size $\omega_x>0$ (respectively, $\omega_x<0$) is a commitment by its owner to sell (respectively, buy) up to $\left|\omega_x\right|$ units of the asset at a price no less than (respectively, no greater than) $p_x$.

Whenever an institution submits a buy (respectively, sell) order $x$, an LOB's trade-matching algorithm checks whether it is possible for $x$ to \emph{match to} an active sell (respectively, buy) order $y$ such that $p_y \leq p_x$ (respectively, $p_y \geq p_x$). If so, the matching occurs immediately and the owners of the relevant orders agree a trade for the specified amount at the specified price. If not, then $x$ becomes \emph{active,} and it remains active until either it matches to an incoming sell (respectively, buy) order, or it is \emph{cancelled}.

Orders that result in an immediate matching upon arrival are called \emph{market orders}. Orders that do not --- instead becoming active orders --- are called \emph{limit orders}.\footnote{Some platforms allow other order types (such as fill-or-kill, stop-loss, or peg orders \citep{HotspotGUIUserGuide}), but it is always possible to decompose the resulting order flow into limit and/or market orders. Therefore, we study LOBs in terms of these simple building blocks.} The \emph{LOB} $\mathcal{L}(t)$ is the set of all active orders for a given asset on a given platform at a given time $t$. For a detailed introduction to LOBs, see \citet{Gould:2013limit}.

At a given time $t$, the \emph{bid price} $b(t)$ is the highest stated price among active buy orders,\begin{equation}b(t):=\max_{\left\{x \in \mathcal{L}(t)|\omega_x<0\right\} } p_x,\end{equation}and the \emph{ask price} $a(t)$ is the lowest stated price among active sell orders,\begin{equation}a(t):=\min_{\left\{x \in \mathcal{L}(t)|\omega_x>0\right\} }p_x.\end{equation}The bid price and ask price are collectively called the \emph{best quotes}. The \emph{bid-ask spread at time $t$} is \begin{equation}s(t):=a(t)-b(t).\end{equation}The \emph{mid price at time $t$} is\begin{equation}m(t):=\frac{a(t)+b(t)}{2}.\end{equation}We say that a price $p$ is \emph{on the buy side of $\mathcal{L}(t)$} if $p\leq b(t)$, \emph{on the sell side of $\mathcal{L}(t)$} if $p\geq a(t)$, or \emph{inside the bid--ask spread} if $b(t)<p<a(t)$.

\subsection{Order Queues in a Limit Order Book}

LOBs implement two \emph{resolution parameters:} the \emph{tick size} $\pi>0$, which specifies the smallest permissible price interval between different orders, and the \emph{lot size} $\sigma>0$, which specifies the smallest amount of the asset that can be traded. All orders must arrive with a price that is a positive integer multiple of $\pi$ and a size that is an integer multiple of $\sigma$. For example, if $\pi = \$0.01$, then the largest permissible order price that is strictly less than $\$1.00$ is $\$0.99$. Similarly, if $\sigma = 10$ shares, the the smallest permissible order size that is strictly greater than $100$ shares is $110$ shares.

\begin{figure}
\centering
\includegraphics[width=8 cm]{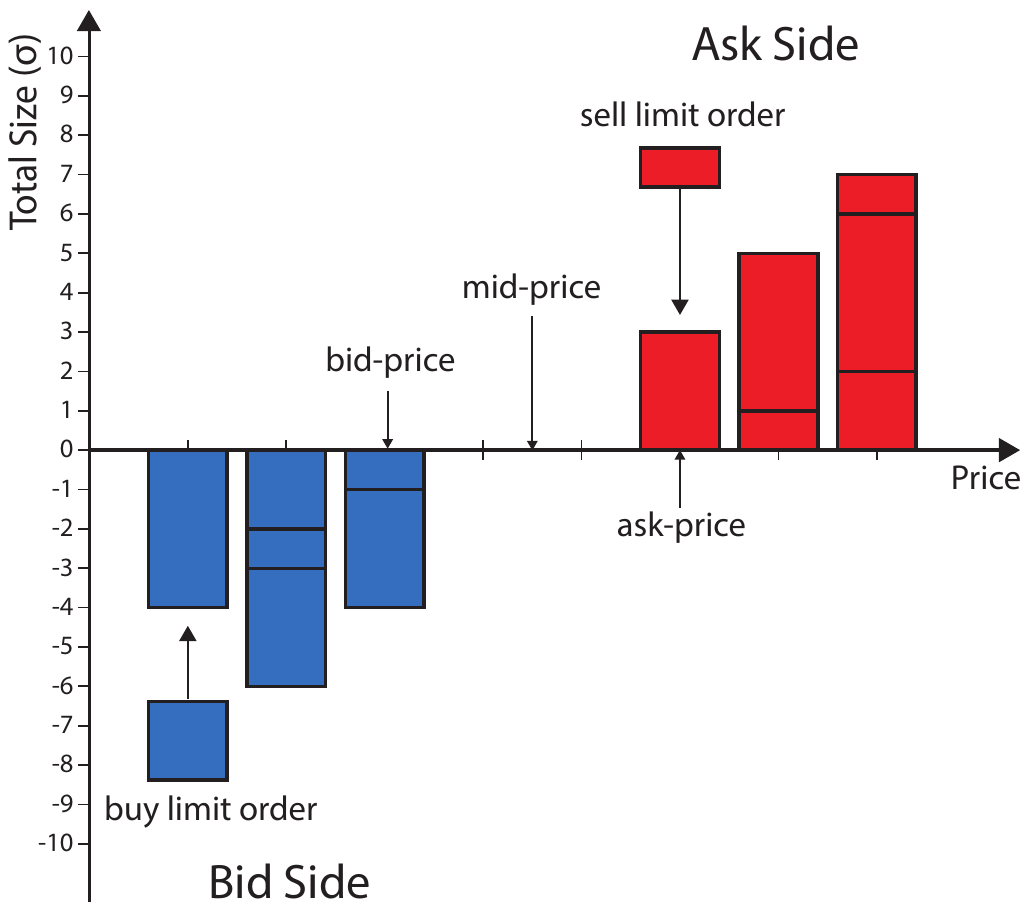}
\caption{Schematic of an LOB. The horizontal lines within the blocks at each price level denote how the total size at that price is composed of different active orders.}
\label{schematiclob}
\end{figure}

Because the tick size is strictly positive, the price axis of an LOB is a one-dimensional lattice, whose points correspond to positive integer multiples of $\pi$. An LOB can therefore be regarded as a set of queues, each of which consists of active buy or sell orders at a specified price (see Figure \ref{schematiclob}). At a given price $p$ and time $t$, the total size of active buy orders (i.e., the length of the queue of buy limit orders) is given by\begin{equation}n^b(p,t):=\sum_{\left\{x \in \mathcal{L}(t)|\omega_x<0, p_x=p\right\}} \left| \omega_x \right|\end{equation}and the total size of active sell orders (i.e., the length of the queue of sell limit orders) is given by\begin{equation}n^a(p,t):=\sum_{\left\{x \in \mathcal{L}(t)|\omega_x>0, p_x=p\right\}} \omega_x.\end{equation}

In an LOB, the value of $b(t)$ increases whenever a new buy limit order arrives inside the bid--ask spread, and decreases whenever the total volume of buy limit orders at $b(t)$ depletes to 0 (which occurs when all buy limit orders at $b(t)$ either match to an incoming sell market order or are cancelled by their owners). Similarly, the value of $a(t)$ decreases whenever a new sell limit order arrives inside the bid--ask spread, and increases whenever the total volume of sell limit orders at $a(t)$ depletes to 0. The value of $m(t)$ increases (respectively, decreases) whenever either either of $b(t)$ or $a(t)$ increases (respectively, decreases).

\section{Price Prediction and Queue Imbalance in an LOB}\label{sec:priceprediction}

During the past 20 years, many empirical and theoretical studies have sought to establish links between the state of an LOB and the subsequent price changes that occur within it. The literature on this topic is vast, and spans many different disciplines, including economics, physics, mathematics, statistics, and psychology. For a recent survey of work in this field, see \citet{Gould:2013limit}. In this section, we review a selection of publications most relevant to our work, and introduce the notion of queue imbalance, which forms the basis for our empirical study.

\subsection{Zero-Intelligence Models of LOB State}

Early studies of the links between queue dynamics and price formation in an LOB typically assumed that order flows were governed by simple, independent stochastic processes. Several authors postulated models of LOB state to formulate predictions of future price movements. Models of this type are often called ``zero-intelligence'' LOB models, because they are not motivated by rational traders who seek to achieve specified trading goals. \citet{Smith:2003statistical} introduced a zero-intelligence model in which limit order arrivals, market order arrivals, and cancellations all occur as mutually independent Poisson processes with fixed rate parameters. \citet{Cont:2010stochastic} extended this model by allowing the rates of limit order arrivals and cancellations to vary across prices. \citet{Huang:2015simulating} studied a model in which the order arrival rates depend on the lengths of order queues in $\mathcal{L}(t)$. \citet{Mike:2008empirical} considered a zero-intelligence framework in which order flows also exhibit long-range autocorrelations, in agreement with empirical data.

Despite the apparent simplicity of these approaches, zero-intelligence LOB models suffer from several important drawbacks. First, the LOB dynamics that emerge from their interacting order flows are often extremely complex. Consequently, most analysis of zero-intelligence LOB models relies heavily on extensive simulations, rather than analytical treatment. Second, although they make reasonably good predictions of the long-run statistical properties of LOB state (see, e.g., \citet{Farmer:2005predictive}), zero-intelligence LOB models typically produce poor predictions of price movements. Third, because such models ignore the strategies implemented by real traders, they fail to capture many important correlations and feedback loops between different order flows. This weakness further harms their ability to make good predictions of future price movements in real markets.

\subsection{Simplifying the State Space}

Motivated by these weaknesses, several authors have proposed models that seek to produce more realistic LOB dynamics on a simplified LOB state space. \citet{Cont:2013price} introduced a model in which $n^b(b(t),t)$ and $n^a(a(t),t)$ are assumed to be governed by independent diffusion processes. By studying this model in the hydrodynamic limit, in which stochastic fluctuations are dominated by deterministic flows, the authors obtained analytical expressions for several quantities of interest, such as the distribution of times between price changes, the distribution and autocorrelation of price changes, and the probability of a mid-price movement in a given direction, given $n^b(b_t,t)$ and $n^a(a_t,t)$.

Despite the appealing nature of these results, \citet{Gareche:2013fokker} presented empirical results to illustrate that the dynamics of $n^b(b_t,t)$ and $n^a(a_t,t)$ are strongly influenced by a coupling between the two queues, and thereby brought into question Cont and de Larrard's assumption of independence between the dynamics of $n^b(b_t,t)$ and $n^a(a_t,t)$. \citet{Avellaneda:2011forecasting} studied a similar model in which $n^b(b_t,t)$ and $n^a(a_t,t)$ are also governed by diffusion processes, but with a specified correlation $\rho$. The authors solved their model to deduce a simple, closed-form predictor for the direction of the next mid-price movement, given $n^b(b_t,t)$, $n^a(a_t,t)$, and $\rho$. Although their assumption that $n^b(b_t,t)$ and $n^a(a_t,t)$ are related by a simple correlation parameter is clearly a simplification of reality, the authors argued that their model produces reasonably good predictions of price movements in real LOBs.

\subsection{Queue Imbalance in an LOB}\label{subsec:imbalance}

In a recent publication, \citet{Yang:2015reduced} argued that the property of the best bid and ask queues most useful for price prediction is not their lengths, but rather their imbalance. Specifically, at a given time $t$, let\begin{equation}\label{eq:I}I(t):= \frac{n^b(b_t,t)-n^a(a_t,t)}{n^b(b_t,t)+n^a(a_t,t)}\end{equation}denote the \emph{queue imbalance at time $t$}. The quantity $I$ measures the (normalized) difference between $n^b(b_t,t)$ and $n^a(a_t,t)$, and thereby provides a quantitative assessment of the relative strengths of buying and selling pressure in an LOB.\footnote{\citet{Yang:2015reduced} define the queue imbalance using only $n^b(b_t,t)$ on the numerator. This produces a linear rescaling of our definition in Equation (\ref{eq:I}), such that $I \in \left[0,1\right]$. We choose to use $n^b(b_t,t)-n^a(a_t,t)$ as the numerator because it instead produces an imbalance on the interval $\left[-1,1\right]$, and thereby simplifies sign conventions.}

If $I \approx 0$ (which occurs when $n^b(b_t,t)$ and $n^a(a_t,t)$ are approximately equal), then the buying and selling pressures are approximately balanced. If $I>0$ (which occurs when $n^b(b_t,t)>n^a(b_t,t)$), then the bid queue is longer than the ask queue, which suggests that there is a net positive buying pressure in the LOB. Conversely, if $I<0$ (which occurs when $n^b(b_t,t)<n^a(b_t,t)$), then the ask queue is longer than the bid queue, which suggests that there is a net positive selling pressure in the LOB. Values of $I$ close to $1$ suggest a very strong buying pressure, and values of $I$ close to $-1$ suggest a very strong selling pressure.

Empirical study of how the queue imbalance in an LOB affects future price changes dates back almost a decade, to when \citet{Cao:2009information} studied how the value of $I$ at a specified depth inside the LOB (i.e., not only at the best quotes) influenced mid-price returns during the subsequent 5-minute interval. \citet{Stoikov:2015reducing} noted that the optimal execution of a single order over a short time horizon depends on the queue imbalance in an LOB. \citet{Cartea:2015enhancing} studied queue imbalance at the trade-by-trade level on Nasdaq, and noted that the rate of buy (respectively, sell) market order arrivals and the probability of observing an upward (respectively, downward) price movement both increase considerably when $I$ is strongly positive (respectively, strongly negative). Based on their analysis, they designed a trading algorithm whose objective is to place both buy and sell limit orders and thereby generate profits from round-trip trades.

Similarly to these previous publications, the aim of the present paper is to investigate whether queue imbalance $I$ influences the future price dynamics in an LOB. Specifically, we seek to complement the analyses of queue imbalance in these papers with a more formal, quantitative analysis of whether the queue imbalance $I$ in an LOB provides significant predictive power for the direction of the next mid-price movement. In contrast to the previous work in this area, we do not restrict our attention to large-tick stocks, but rather study the predictive power of queue imbalance for a selection of both small-tick and large-tick stocks. By comparing our results from these different stocks, we seek to illuminate how the underlying market microstructure underpins the usefulness of $I$ as a predictor of future price movements.

\section{Data}\label{sec:data}

The data that we study originates from the LOBSTER database, which provides an event-by-event description of the temporal evolution of the LOB for each stock listed on Nasdaq.\footnote{For a detailed introduction to LOBSTER, see \url{http://LOBSTER.wiwi.hu-berlin.de}.} The LOBSTER database contains very detailed information regarding the temporal evolution of the relevant LOBs. However, for the present study we require only the time series of bid prices $b(t)$, ask prices $a(t)$, and queue lengths $n^b(b(t),t)$ and $n^a(a(t),t)$. We derive all of the statistics that we use for our empirical analysis from these 4 simple time series. To produce our empirical results in Section \ref{sec:results}, we study these time series for a selection of 10 liquid stocks during the entire year of 2014.

The Nasdaq platform operates continuous trading from $09$:$30$ to $16$:$00$ on each weekday. Trading does not occur on weekends or public holidays, so we exclude these days from our analysis. We also exclude all activity during the first and last $30$ minutes of each trading day, to ensure that our results are not affected by the abnormal trading behaviour that can occur shortly after the opening auction or shortly before the closing auction. After making these exclusions, we therefore study all trading activity from $10$:$00$ to $15$:$30$ on each of the 252 trading days in 2014.

On the Nasdaq platform, each stock is traded in a separate LOB with price--time priority, with a tick size of $\pi = \$0.01$ (see Section \ref{sec:lobs}). Although this tick size is common to all stocks, the prices of different stocks on Nasdaq vary across several orders of magnitude (from about $\$1$ to more than $\$1000$). Therefore, the \emph{relative tick size} (i.e., the ratio between the stock price and $\pi$) varies considerably across different stocks.

In order to facilitate comparisons between stocks with different relative tick sizes, we first ranked all stocks listed on the Nasdaq exchange according to their total dollar volume of trades during 2014. From this list, we then selected the top 5 entries whose maximal trade price was below $\$50.00$, and the top 5 entries whose minimal trade price was above $\$100.00$ (see Table~\ref{tab:summaries}). We call the first group of stocks \emph{large-tick stocks} because their low price makes their relative tick size large, and we call the second group of stocks \emph{small-tick stocks} because their high price makes their relative tick size small. We also repeated our calculations for several other stocks whose prices fell between these two thresholds. We found that some such stocks behaved similarly to the large-tick stocks in our sample, whereas other behaved similarly to the small-tick stocks in our sample. In order to illustrate the clear separation between the results for different types of stocks, we present only our results for the large-tick and small-tick stocks in our sample, and not for the other stocks with intermediate tick size. Table \ref{tab:summaries} lists the names of the stocks that we choose in this way, along with several summary statistics that describe their aggregate market activity.

\begin{sidewaystable}
\begin{center}
\begin{tabular}{|l|c|c|c|c|c|c|c|c|}
\hline
 & $V_{MO}$ ($\$$bn) & $V_{LO}$ ($\$$bn) & $p_{\min}$ ($\$$) & $p_{\max}$ ($\$$) & $\left\langle{n^b(b(t),t)}\right\rangle$ & $\left\langle{n^a(a(t),t)}\right\rangle$ & $\left\langle{s(t)}\right\rangle$ ($\$$) \\
\hline
MSFT & $48.22$ & $782.04$ & $34.97$ & $49.94$ & $6206.7$ & $6286.8$ & $0.013$ \\ 
INTC & $31.22$ & $462.71$ & $23.50$ & $37.84$ & $9597.5$ & $9879.5$ & $0.013$ \\ 
MU & $28.72$ & $341.83$ & $20.65$ & $36.50$ & $3802.9$ & $3922.8$ & $0.014$ \\ 
CSCO & $24.99$ & $448.87$ & $21.29$ & $28.59$ & $17114.9$ & $18128.0$ & $0.012$ \\ 
ORCL & $16.00$ & $254.02$ & $35.65$ & $46.70$ & $3113.7$ & $3244.9$ & $0.015$ \\ 
\hline
GOOG & $62.05$ & $265.70$ & $498.32$ & $1228.88$ & $133.4$ & $134.9$ & $0.360$ \\ 
AMZN & $57.02$ & $146.99$ & $285.05$ & $407.80$ & $158.1$ & $155.6$ & $0.195$ \\ 
TSLA & $51.37$ & $111.60$ & $136.90$ & $291.40$ & $204.4$ & $203.2$ & $0.212$ \\ 
PCLN & $43.24$ & $304.14$ & $1017.07$ & $1375.18$ & $80.3$ & $80.2$ & $1.111$ \\ 
NFLX & $40.69$ & $135.79$ & $299.50$ & $488.98$ & $120.3$ & $120.0$ & $0.352$ \\ 
\hline
\end{tabular}
\caption{Summary statistics describing trading activity for the (top panel) large-tick and (bottom panel) small-tick stocks in our sample, which describes LOB activity on Nasdaq during 2014. The large-tick stocks are Microsoft (MSFT), Intel (INTC), Micron Technology (MU), Cisco Systems (CSCO), and Oracle (ORCL). The small-tick stocks are Google (GOOG), Amazon (AMZN), Tesla Motors (TSLA), Priceline (PCLN), and Netflix (NFLX). For each stock, the columns list the total size of market order arrivals $V_{MO}$ (measured in billions of $\$$), the total size of limit order arrivals at the best quotes $V_{LO}$ (measured in billions of $\$$), the minimum trade price $p_{\min}$ (measured in $\$$), the maximum trade price $p_{\max}$ (measured in $\$$), the mean size of the bid queue $\left\langle{n^b(b(t),t)}\right\rangle$ (measured in number of shares), the mean size of the ask queue $\left\langle{n^a(a(t),t)}\right\rangle$ (measured in number of shares), and the mean bid--ask spread $\left\langle{s(t)}\right\rangle$ (measured in $\$$).}
\label{tab:summaries}
\end{center}
\end{sidewaystable}

The LOBSTER data has many important benefits that make it particularly suitable for our study. First, the data is recorded directly by the Nasdaq servers. Therefore, we avoid the many difficulties associated with data sets that are recorded by third-party providers, such as misaligned time stamps or incorrectly ordered events. Second, the data is fully self-consistent, in the sense that it does not report any activities or updates that would violate the standard rules of LOB trading. By contrast, many other LOB data sets suffer from recording errors that can constitute a considerable source of noise when performing detailed analysis. Third, each limit order described in the data constitutes a firm commitment to trade. Therefore, our results reflect the market dynamics for real trading opportunities, not ``indicative'' declarations of possible intent.

The LOBSTER database describes all LOB activity that occurs on Nasdaq, but does not provide any information regarding order flow for the same assets on different platforms. To minimize the possible impact on our results, we restrict our attention to stocks for which Nasdaq is the primary trading venue and therefore captures the majority of order flow. Our results enable us to identify several robust statistical regularities linking queue imbalance to the directions of mid-price movements, which is precisely the aim of our study. We therefore do not regard this feature of the LOBSTER data to be a serious limitation for the present study.

\section{Methodology}\label{sec:methodology}

\subsection{Sample Construction}

For each stock and each trading day in our sample, we first create an ordered set $T$ of times at which the mid price changes,\begin{equation}\label{eq:mchange}T = \left\{t \ \middle| \ m(t) \neq \lim_{\varepsilon \downarrow 0}m(t-\varepsilon)\right\}.\end{equation}Let $t_1<t_2< \ldots< t_N$ denote the times in $T$, and let $t_0$ denote the time of the first LOB event for the given stock on the given day (which, for our data, occurs at or after $10$:$00$ --- see Section \ref{sec:data}).

For each time $t_i \in T$, we calculate an indicator variable $y_i$ to describe whether or not the mid-price movement at $t_i$ was upwards,\begin{equation}\label{eq:yi}y_i:=\left\{\begin{array}{ll}
1, &\text{ if }m(t_i)>m(t_{i-1}),\\
0, &\text{ if }m(t_i)<m(t_{i-1}).\end{array}\right.\end{equation}We choose to study price changes via this simple indicator variable, rather than studying the signed change in mid price, because the magnitude of such price changes are determined not only by $n^b(b(t),t)$ and $n^a(a(t),t)$, but also by the prices and lengths of order queues deeper into the LOB. We therefore restrict our attention to the direction of mid-price movements, not their size.

For each time $t_i \in T$, we choose a time $\tilde{t}_i$ uniformly at random in the open interval $(t_{i-1},t_i)$ and sample the imbalance $I(\tilde{t}_i)$ at this time. To ease exposition, we introduce the notation\begin{equation}I_i=I(\tilde{t}_i).\end{equation}In this paper, we seek to assess the predictive power of $I_i$ for forecasting $y_i$. There are many other possible choices for when to sample the imbalance to perform predictions (such as sampling $I$ immediately after $t_{i-1}$ or immediately before $t_i$), but we restrict our attention to the case of sampling $\tilde{t}_i$ uniformly at random. We return to the discussion of possible alternative approaches in Section \ref{sec:conclusions}.

For all stocks in our sample, both the number and temporal spacing between LOB events that affect $n^b(b(t),t)$ and $n^a(a(t),t)$ (i.e., market order arrivals and limit order arrivals and cancellations at the best quotes) varies considerably across different time intervals $(t_{i-1},t_i)$. In some cases, the arrivals of such events are highly clustered, in the sense that the time intervals contain some periods with no updates to $n^b(b(t),t)$ and $n^a(a(t),t)$ and other periods with many updates to $n^b(b(t),t)$ and $n^a(a(t),t)$. In order to understand whether this event clustering strongly influences our results, we also repeated all of our calculations when constructing our random sample in \emph{event time}. To do so, for each time $t_i \in T$, we first construct a list of the times that either $n^b(b(t),t)$ or $n^a(a(t),t)$ changed, during the time interval $(t_{i-1},t_i)$. We then choose our sampling time $\tilde{t}_i$ uniformly at random from this (discrete) set of event times. We found that all of our empirical results when choosing $\tilde{t}_i$ in this way were qualitatively the same as those that we report throughout the paper, for which we choose $\tilde{t}_i$ uniformly at random in the open interval $(t_{i-1},t_i)$.

\subsection{In-Sample and Out-of-Sample Data}

The number of mid-price changes that occur in a single trading day varies considerably both across different stocks and across different days. To ensure that our sample contains the same number of data points for each stock each day, we therefore draw a random subsample with a fixed size among the $N$ possible choices in the set $T$. For the results that we show in Section \ref{sec:results}, we use a random subsample size\footnote{We also repeated all of our calculations with a variety of subsample sizes ranging from $50$ to $1000$, and we found that our results were qualitatively the same in each case.} of 100. For each stock, we then aggregate the subsamples from each of the 252 different trading days to produce an aggregated data set of 25200 data points.

A key contribution of the present work is assessing the strength of predictive power provided by $I$. To avoid the possible dangers of data snooping when performing this analysis, we randomly partition each stock's aggregated data set into two disjoint subsets: a training set, which contains $80\%$ of the data (i.e., 20160 data points), and a testing set, which contains $20\%$ of the data (i.e., 5040 data points). We perform the fits of our logistic regressions and local logistic regressions using the training set (i.e., ``in sample''), then evaluate the predictive power of these fits using the testing set (i.e., ``out of sample'').

\subsection{Formulating Predictions}\label{subsec:logistic}

The aim of the present study is to assess the predictive power of $I_i$ for forecasting $y_i$. We first consider this question in the context of a simple binary classifier, which seeks to predict whether, for a given queue imbalance $I_i$, the value of $y_i$ will be $0$ or $1$ (i.e., whether the direction of the next mid-price movement will be upwards or downwards). To perform this binary classification, we seek to estimate a function $\hat{y}$ that maps queue imbalance onto some subset of $\mathbb{R}$, and a threshold value $y^{*}\in \mathbb{R}$, such that:
\begin{itemize}
\item if $\hat{y}(I_i)>y^{*}$, then we predict $y_i$ to equal $1$,
\item if $\hat{y}(I_i)<y^{*}$, then we predict $y_i$ to equal $0$,
\item if $\hat{y}(I_i)=y^{*}$, then we predict $y_i$ to equal either $0$ or $1$, each with probability $1/2$.
\end{itemize}

We then consider this question in the context of a probabilistic classifier, which, for a given queue imbalance $I_i$, seeks to predict the probability that $y_i=1$. We note that if we choose the function $y_i$ for our binary classifier such that\begin{equation}\label{eq:yhay}\hat{y}:\left(-1,1\right) \rightarrow \left[0,1\right],\end{equation}and if we interpret $\hat{y}$ as\begin{equation}\hat{y}_i:=\mathbb{P}\left(y_i=1 \middle| I_i \right),\end{equation}then we can use the same function $\hat{y}$ to perform both binary classification and probabilistic prediction.

Consider a queue imbalance $I_{i'}$ chosen uniformly at random among all observations for which $y_i=1$ and another queue imbalance $I_{j'}$ chosen uniformly at random among all observations for which $y_i=0$. If $I_i$ provides predictive power to perform binary classification, then the resulting values of $\hat{y}$ will satisfy\begin{equation}\mathbb{P}(\hat{y}_{i'}>\hat{y}_{j'})>1/2.\end{equation}If, however, $I_i$ provides no predictive power to perform binary classification, then the resulting values of $\hat{y}$ will satisfy\begin{equation}\mathbb{P}(\hat{y}_{i'}>\hat{y}_{j'})=\mathbb{P}(\hat{y}_{j'}>\hat{y}_{i'})=1/2.\end{equation}Similarly, if $I_i$ provides predictive power to perform probabilistic classification, then the resulting values of $\hat{y}$ will satisfy\begin{equation}\mathbb{P}(\hat{y}_{i'}>1/2)>1/2 \text{ and }\mathbb{P}(\hat{y}_{j'}>1/2)<1/2.\end{equation}If, however, $I_i$ provides no predictive power to perform probabilistic classification, then the resulting values of $\hat{y}$ will satisfy\begin{equation}\mathbb{P}(\hat{y}_{i'}>1/2)=1/2 \text{ and }\mathbb{P}(\hat{y}_{j'}>1/2)=1/2.\end{equation}

To formulate our estimate of the function $\hat{y}$, we perform a logistic regression of $y_i$ onto $I_i$. Specifically, we use the data in our training set to calculate maximum likelihood estimates of the coefficients $x_0$ and $x_1$ in the relationship\begin{equation}\label{eq:logistic}\hat{y}(I)=\frac{1}{1+e^{-(x_0+Ix_1)}}.\end{equation}For a detailed introduction to logistic regression, see \citet{Hosmer:2004applied,Mccullagh:1989generalized}.

\subsection{Assessing Predictions}\label{subsec:assessing}

To assess the predictive power of our logistic regressions for performing binary and probabilistic classification, we compare their output to that of a simple null model in which we assume that $I$ provides no useful information for predicting the direction of mid-price movements, such that\begin{equation}\label{eq:nullmodel}\hat{y}(I) = 1/2\text{ for all }I.\end{equation}In words, our null model predicts that the probability of an upward price movement is always $1/2$, irrespective of the queue imbalance.

To assess the predictive power of our fits for performing binary classification, we calculate the out-of-sample receiver operating characteristic (ROC) curves and the corresponding area-under-ROC-curve statistics. The area under the ROC curve quantifies how successfully the logistic regression fits classify cases that result in a price move of a given direction. More precisely, for a given queue imbalance $I_{i'}$ chosen uniformly at random among all observations for which $y_i=1$ and another queue imbalance $I_{j'}$ chosen uniformly at random among all observations for which $y_i=0$, the area under the ROC curve is equal to the probability that the resulting values of $\hat{y}$ will satisfy $\hat{y}_{i'}>\hat{y}_{j'}$. For a detailed introduction to ROC curves, see \citet{Bradley:1997use} and \citet{Hanley:1982meaning}.

To assess the predictive power of our fitted logistic regressions for performing probabilistic classification, we use our function $\hat{y}$ to make out-of-sample forecasts $\hat{y}_i$ for each $I_i$ in the testing set, and calculate the corresponding residuals\begin{equation}\label{eq:ri}r_i:=\hat{y}_i-y_i.\end{equation}We then calculate the mean square residual across all observations in the testing set. For the null model, $\hat{y}_i=0.5$ for all $i$, so $r_i$ is given by\begin{equation}\label{eq:rinull}r_i = \left\{\begin{array}{ll}
-1/2, &\text{ if }y_i=1,\\
1/2, &\text{ if }y_i=0\end{array}\right.\end{equation}and the mean squared residual is exactly $1/4$.

\subsection{Local Logistic Regression}\label{subsec:locallogit}

After performing our logistic regression fits, we also consider an alternative estimation of the function $\hat{y}$ by performing a local logistic regression of $y_i$ onto $I_i$. A local logistic regression is a semi-parametric estimation method that fits a separate logistic regression at each point in the domain. Specifically, for a given imbalance $I$, the local logistic regression estimator of $\hat{y}(I)$ is obtained by performing a standard logistic regression at $I$, but weighting the input observations according to their distance from $I$. For a detailed introduction to local logistic regression, see \citet{Loader:2006local}. When performing our local logistic regression fits, we use a standard tricube weight function with a nearest-neighbour bandwidth parameter, whose value we choose by performing a 5-fold cross validation in the training set. For a detailed discussion of parameter estimation via cross validation, see \citet{Hastie:2009elements}.

\section{Results}\label{sec:results}

We now present our main empirical results. In Section~\ref{subsec:resultsdistribution}, we calculate the distribution of queue imbalances that we observe in our sample. We perform our logistic regression fits in Section~\ref{subsec:resultslogistic} and assess their out-of-sample performance in Section~\ref{subsec:resultsoos}. In Section~\ref{subsec:resultslocallogistic}, we perform our local logistic regression fits and analyze the resulting curves.

\subsection{Distribution of $I$}\label{subsec:resultsdistribution}

\begin{figure}[!htbp]
\centering
\includegraphics[width=0.9\textwidth]{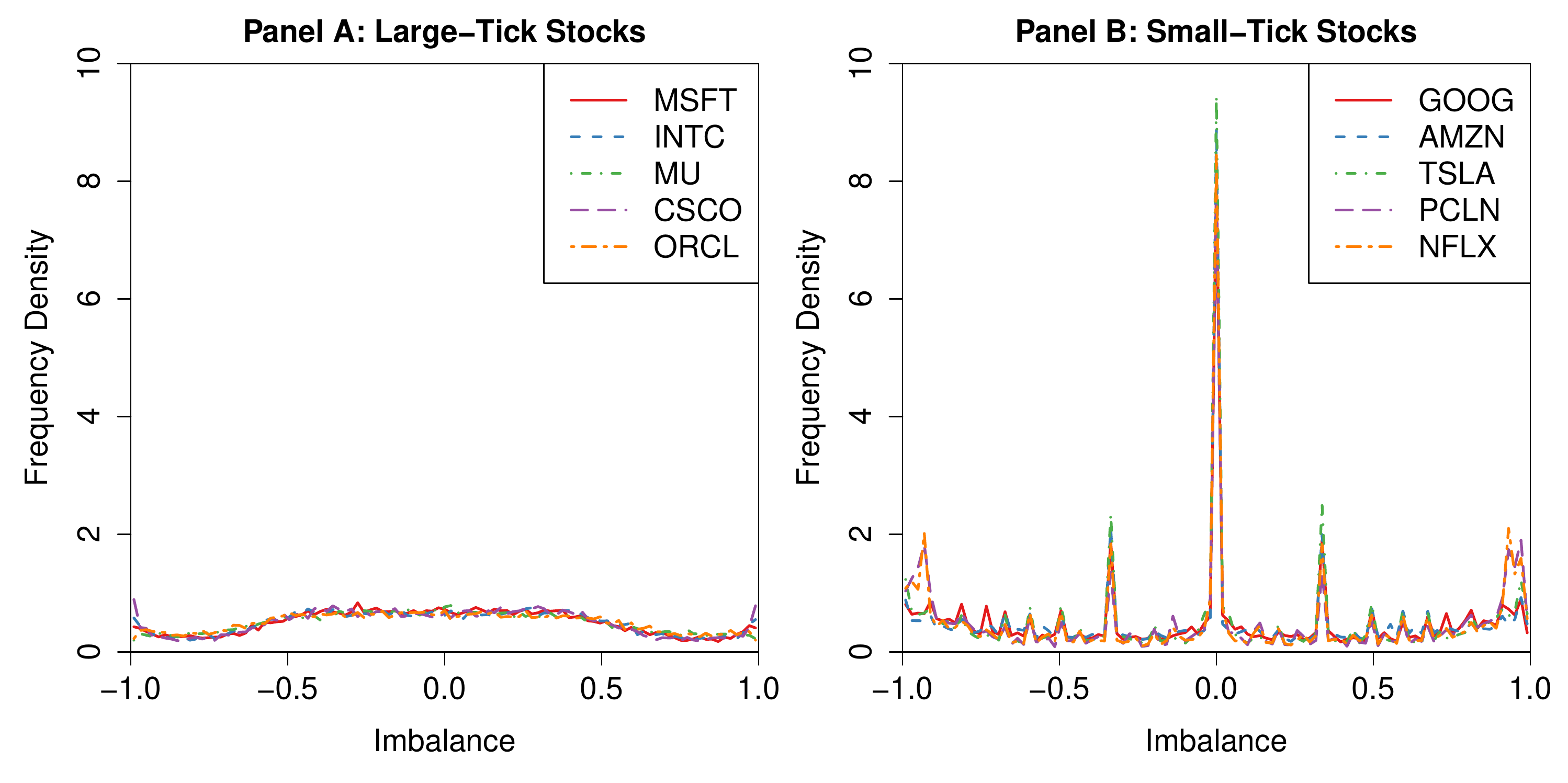}
\caption{Histograms of $I$ for each of the 10 stocks in our sample. The left panel shows the results for large-tick stocks and the right panel shows the results for small-tick stocks.}
\label{fig:randHist}
\end{figure}

To help understand the distribution of queue imbalances that occur in our sample, we first calculate histograms of $I_i$ for each of the 10 stocks. We plot these histograms in Figure~\ref{fig:randHist}.

For large-tick stocks, it is common to observe a wide range of imbalances between about $-0.5$ and about $0.5$. The distribution of $I$ decays gradually outside of this range. For some large-tick stocks, this decrease is approximately monotonic, up to statistical noise. For others, such as CSCO, there is another small local maximum close to $I = \pm 1$.

\begin{figure}[!htbp]
\centering
\includegraphics[width=0.9\textwidth]{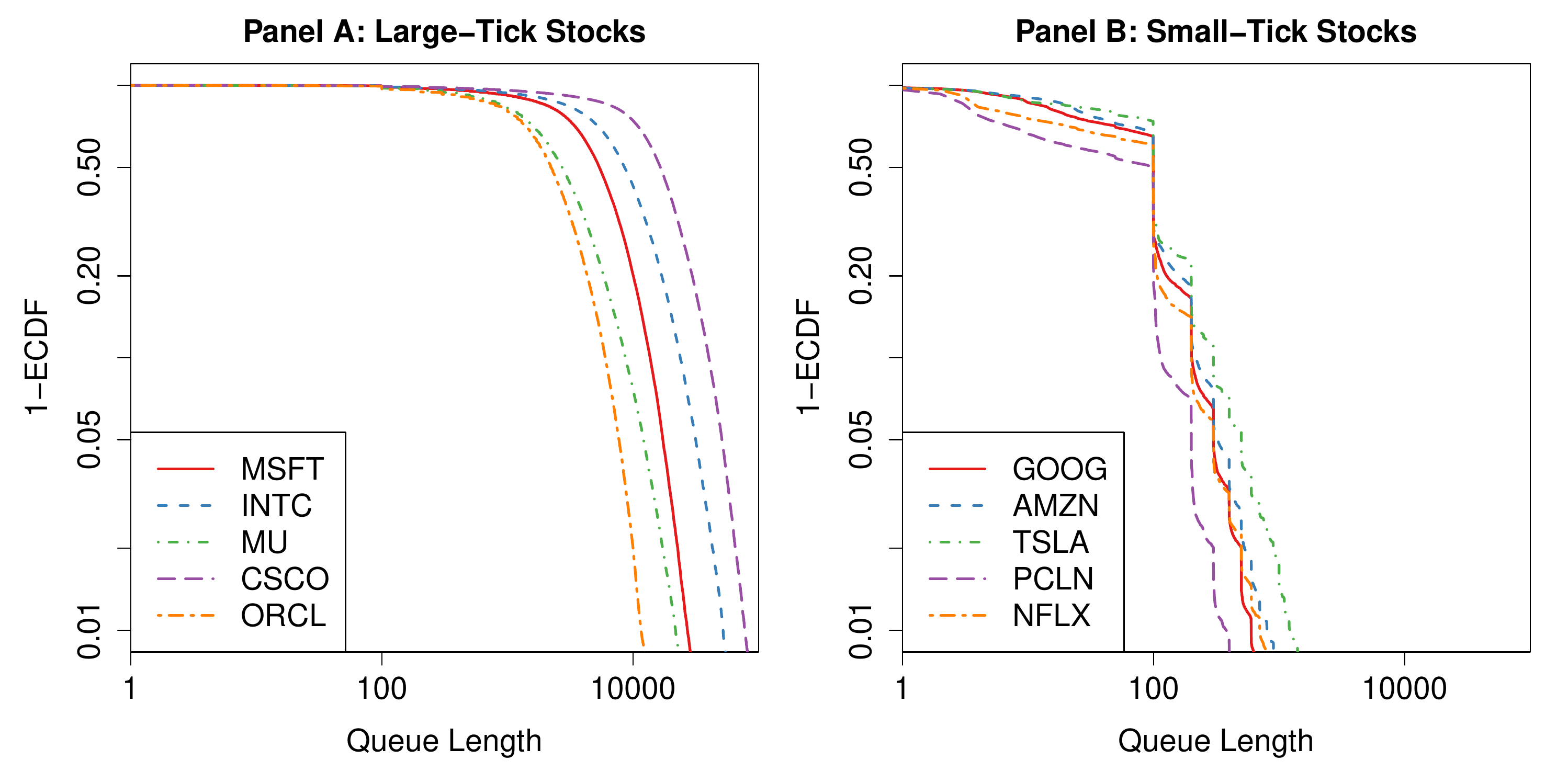}
\caption{Empirical cumulative density functions (ECDFs) of the best-quote queue lengths $n^b(b(t),t)$ and $n^a(a(t),t)$. In order to illustrate the tail behaviour, the plots show the survivor functions (i.e., $1-\text{ECDF}$) in doubly logarithmic coordinates.}
\label{fig:randQueueLength}
\end{figure}

For small-tick stocks, the story is less straightforward due to a strong round-number effect that appears at some values of $I$. Specifically, there exist several queue imbalances that occur much more commonly than do their neighbouring values. This effect is particularly prominent at $I=0$ and $I=\pm1/3$, but is also visible at several other round-number values in the domain of $I$. To illustrate why these round-number effects appear strongly for small-tick stocks, but not for large-tick stocks, we also calculate the empirical cumulative density functions (ECDFs) of the best-quote queue lengths $n^b(b(t),t)$ and $n^a(a(t),t)$ (see Figure \ref{fig:randQueueLength}). For large-tick stocks, the curves decay smoothly across the whole domain, which indicates that it is common to observe queues of a wide range of different lengths. For small-tick stocks, by contrast, the ECDFs contain large jumps at some round numbers, such as $100$ and $200$. This implies that it is much more common to observe queues of these round-number lengths than it is to observe queues of other lengths. These round-number effects for small-tick stocks subsequently manifest in the histograms in Figure \ref{fig:randHist}. For example, if $n^b(b(t),t)=200$ and $n^a(a(t),t)=100$ (both of which occur commonly), then $I(t)=(200-100)/(200+100)=1/3$.

\subsection{Logistic Regression Fits}\label{subsec:resultslogistic}

\begin{figure}[!htbp]
\centering
\includegraphics[width=0.9\textwidth]{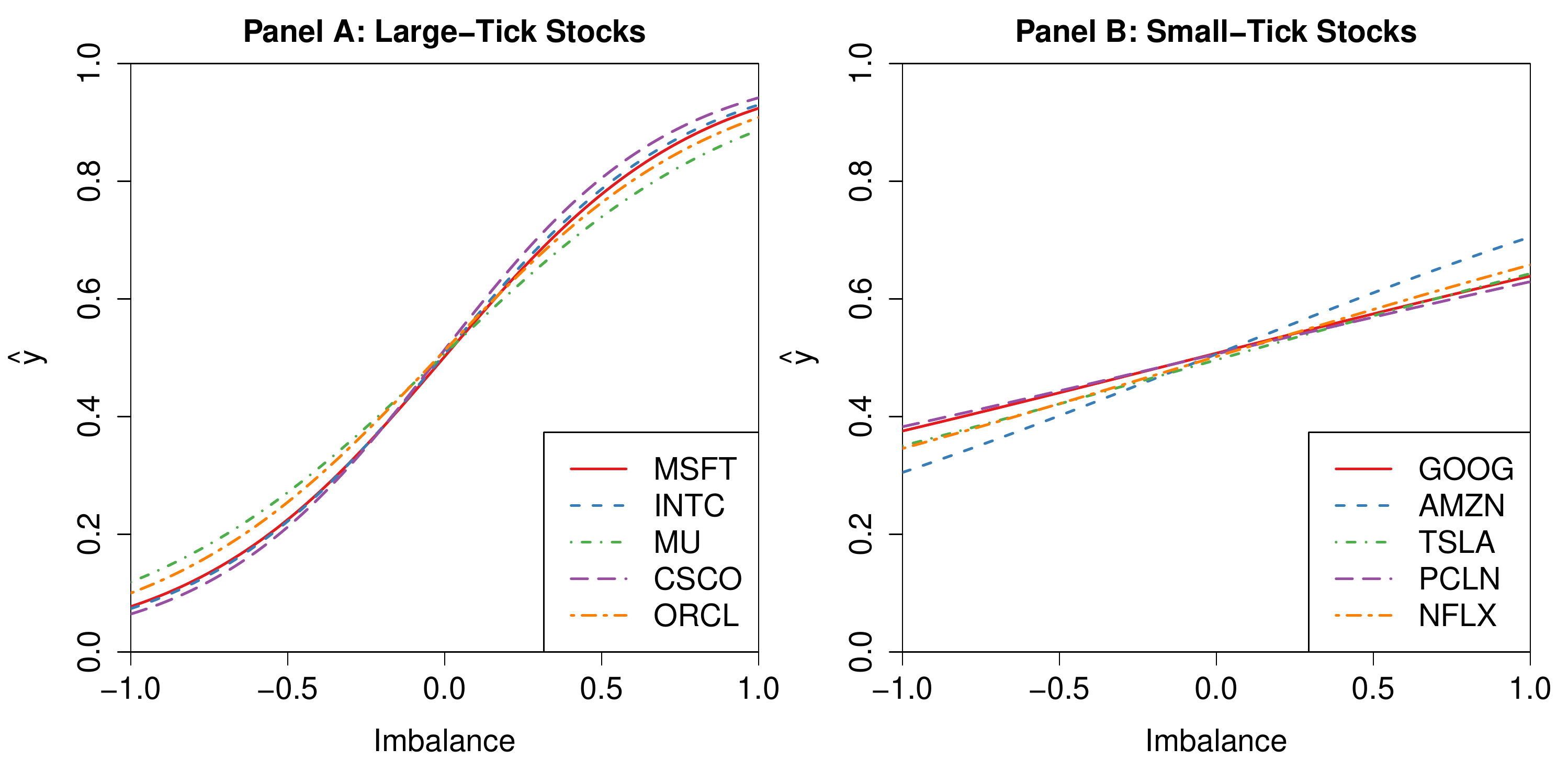}
\caption{Logistic regression fits of $\hat{y}$ versus $I$. The left panel shows the results for large-tick stocks and the right panel shows the results for small-tick stocks.}
\label{fig:randLogit}
\end{figure}

We next perform our logistic regression fits of $\hat{y}$ versus $I$ (see Section \ref{subsec:logistic}). Figure \ref{fig:randLogit} shows the fitted logistic regression curves for each of the stocks in our sample, and Table \ref{tab:randLogit} shows the corresponding maximum likelihood estimates and standard errors of the logistic regression coefficients.

\begin{table}
\begin{center}
\begin{small}
\begin{tabular}{|l|cc|cc|}
\hline
 & \multicolumn{2}{|c|}{$x_0$} & \multicolumn{2}{|c|}{$x_1$} \\
\cline{2-5}
 & Estimate & St. Err. & Estimate & St. Err. \\
\hline
MSFT & $0.01$ & $(0.02)$ & $2.49$ & $(0.04)$ \\ 
INTC & $0.03$ & $(0.02)$ & $2.56$ & $(0.04)$ \\ 
MU & $0.03$ & $(0.02)$ & $2.03$ & $(0.04)$ \\ 
CSCO & $0.06$ & $(0.02)$ & $2.73$ & $(0.04)$ \\ 
ORCL & $0.05$ & $(0.02)$ & $2.25$ & $(0.04)$ \\ 
\hline
GOOG & $0.03$ & $(0.01)$ & $0.54$ & $(0.02)$ \\ 
AMZN & $0.03$ & $(0.01)$ & $0.85$ & $(0.03)$ \\ 
TSLA & $-0.01$ & $(0.01)$ & $0.60$ & $(0.03)$ \\ 
PCLN & $0.03$ & $(0.01)$ & $0.50$ & $(0.02)$ \\ 
NFLX & $0.01$ & $(0.01)$ & $0.65$ & $(0.02)$ \\ 
\hline
\end{tabular}
\caption{Maximum likelihood estimates of the intercept $x_0$ and coefficient $x_1$ in the logistic regression fits of $\hat{y}$ versus $I$. The top panel shows the results for large-tick stocks and the bottom panel shows the results for small-tick stocks. The numbers in parentheses indicate 1 standard error, which we estimate from the corresponding Fisher information matrix. For a full discussion of our logistic regression methodology, see Section \ref{subsec:logistic}.}
\label{tab:randLogit}
\end{small}
\end{center}
\end{table}

Several particularly salient features are apparent from these results. First, the fitted value of $x_0$ is small for all stocks in our sample. This suggests that there is an approximately symmetric behaviour for buy-side and sell-side activity. Specifically, the relationship between upward price movements for a given imbalance $I=k$ is approximately the same as the relationship between downward price movements for a given imbalance $I=-k$, for $k \in \left[0,1\right]$. By continuity of the logistic regression function, it thereby follows that upward and downward price movements are approximately equally likely when $I=0$.

Second, the fitted value of $x_1$ is positive in each case. This implies that the fitted logistic regression line is a monotone increasing function of $I$, and therefore suggests that the larger the queue imbalance, the higher the probability that the next mid-price movement will be upwards.

Third, the fitted values of $x_1$ are much larger for large-tick stocks (for which they vary from about $2$ to about $3$) than for small-tick stocks (for which they vary from about $0.5$ to about $0.8$). These differences in parameter estimates similarly produce substantial differences in the logistic regression fits. For large-tick stocks, the large values of $x_1$ produce substantial curvature in the fitted logistic regression curves (see the left panel of Figure \ref{fig:randLogit}), whereas for small-tick stocks, the small values of $x_1$ produce much flatter, shallower fitted logistic regression curves (see the right panel of Figure \ref{fig:randLogit}). Together, these results suggest that strong imbalances (of either sign) lead to stronger levels of predictability for large-tick stocks than they do for small-tick stocks. For example, the logistic regression curves predict that when $I$ is close to 1, the probability of an upward price move is about $0.8$ to $0.9$ for large-tick stocks, but only about $0.6$ to $0.7$ for small-tick stocks.

We next turn to the question of whether the $x_0$ and $x_1$ coefficients cause a statistically significant impact in the output of the logistic regressions. To address this question, we perform a Wald test for the $x_0$ and $x_1$ coefficients in turn (see the first two columns of Table \ref{tab:randLogitWaldLR}).

\begin{table}
\begin{center}
\begin{small}
\begin{tabular}{|l|c|c|c|}
\hline
 & $x_0$ & $x_1$ & Full Model \\
\hline
MSFT & $0.30$ & $3831.73^{**}$ & $5265.76^{**}$ \\ 
INTC & $2.61$ & $4081.01^{**}$ & $5696.82^{**}$ \\ 
MU & $3.06$ & $3171.89^{**}$ & $3976.12^{**}$ \\ 
CSCO & $11.13^{**}$ & $4261.37^{**}$ & $6205.00^{**}$ \\ 
ORCL & $9.59^{**}$ & $3698.73^{**}$ & $4852.68^{**}$ \\ 
\hline
GOOG & $4.74^{*}$ & $469.55^{**}$ & $483.24^{**}$ \\ 
AMZN & $3.00$ & $1002.24^{**}$ & $1061.45^{**}$ \\ 
TSLA & $1.02$ & $564.90^{**}$ & $583.63^{**}$ \\ 
PCLN & $3.28$ & $502.32^{**}$ & $517.13^{**}$ \\ 
NFLX & $0.34$ & $761.92^{**}$ & $791.63^{**}$ \\ 
\hline
\end{tabular}
\caption{Test statistics for (first two columns) Wald tests for individual coefficients and (third column) likelihood ratio test for the full logistic regression fits of $\hat{y}$ versus $I$. The top panel shows the results for large-tick stocks and the bottom panel shows the results for small-tick stocks. For all tests, the asymptotic distribution of the test statistic is given by a $\chi^2$ distribution with 1 degree of freedom, for which the $95\%$ critical value is $3.84$ and the $99\%$ critical value is $6.63$. Entries marked with an asterisk are statistically significant at the $95\%$ level, and entries marked with a double-asterisk are statistically significant at the $99\%$ level.}
\label{tab:randLogitWaldLR}
\end{small}
\end{center}
\end{table}

At the $95\%$ level, the intercept $x_0$ is not statistically significant for MSFT, INTC, MU, AMZN, TSLA, PCLN, or NFLX, but is statistically significant for CSCO, ORCL, and GOOG. For CSCO and ORCL, $x_0$ is also statistically significant at the $99\%$ level. This result suggests that there is a statistically significant asymmetry between buy-side and sell-side activity for these stocks. We stress, however, that the strength of this asymmetry is very weak (see Table~\ref{tab:randLogit}), and that we are only able to detect it with statistical significance because our sample is large. Therefore, even for these stocks, the behaviour of the fitted logistic regression curve is very close to symmetric about $I=0$.

The coefficient $x_1$ is statistically significant at the $99\%$ level for all 10 stocks in our sample. This implies that the logistic regressions detect a strongly statistically significant relationship between the queue imbalance and the direction of the subsequent mid-price movement.

We next turn to the question of the statistical significance of the full logistic regression fits. To address this question, we perform a likelihood ratio test of the fitted logistic regressions against a nested model that contains only the intercept term, and thereby excludes the possible influence of queue imbalance (see the third column of Table \ref{tab:randLogitWaldLR}). For all 10 stocks in our sample, the results of the likelihood ratio test are statistically significant at the $99\%$ level, which allows us to conclude with high statistical confidence that the relationship between $\hat{y}$ and $I$ illustrated by our logistic regressions is highly statistically significant.

\subsection{Assessing Predictive Power}\label{subsec:resultsoos}

In Section~\ref{subsec:resultslogistic}, we concluded from our logistic regressions that the relationship between the queue imbalance and the direction of the subsequent mid-price movement is highly statistically significant. We now turn to the question of how strongly the estimated logistic regression curve $\hat{y}$ improves the out-of-sample performance of binary and probabilistic classification, in comparison to a simple null model, which assumes that the direction of mid-price changes is uncorrelated with the queue imbalance $I$. Due to the large size of our sample, it could be the case that $\hat{y}$ produces a relatively small increase in predictive power, despite the logistic regression fits being statistically significant.

\begin{figure}[!htbp]
\centering
\includegraphics[width=0.9\textwidth]{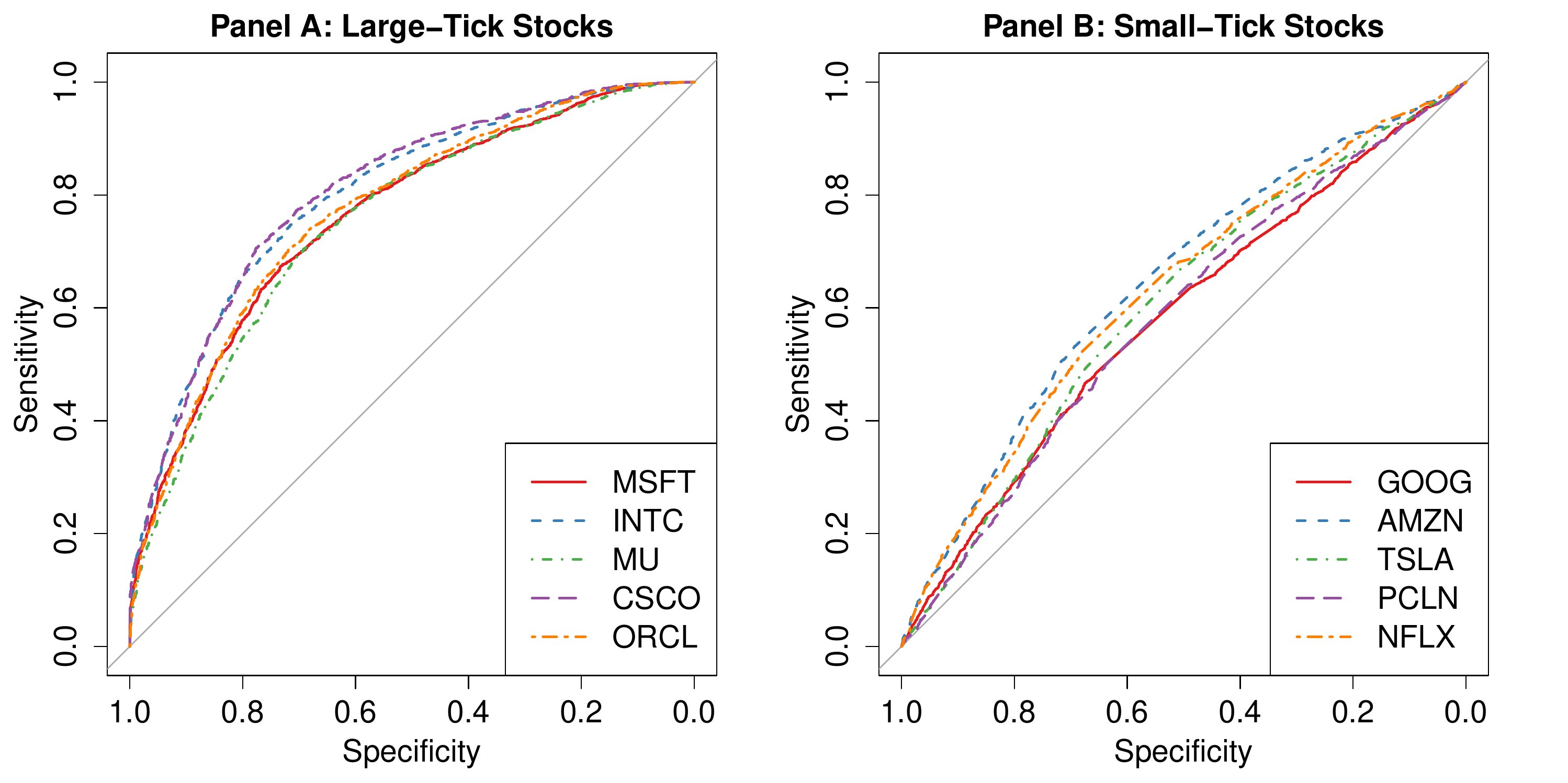}
\caption{Receiver operating characteristic (ROC) curves for the out-of-sample predictive power of a binary classifier, based on the logistic regression fits of $\hat{y}$ versus $I$ from Section~\ref{subsec:resultslogistic}. The left panel shows the results for large-tick stocks and the right panel shows the results for small-tick stocks. The grey line in each plot denotes the expected performance of the null model, which assumes that the probability of an upward price movement is always equal to $1/2$, irrespective of the queue imbalance.}
\label{fig:randLogitROC}
\end{figure}

We first address the out-of-sample performance of our logistic regression fits for performing binary classification. To do so, we calculate the ROC curves (see Section~\ref{subsec:logistic}) for each of our logistic regression fits in Section~\ref{subsec:resultslogistic}. We show these ROC curves, together with the corresponding ROC curve for the null model, in Figure~\ref{fig:randLogitROC}.

In each case, the out-of-sample ROC curves lie above the grey line for all choices of specificity, which implies that the logistic regression fits outperform the out-of-sample predictive power of the null model at all levels of specificity. In Table \ref{tab:randLogitAUC}, we list the area under the ROC curve for each stock (see Section \ref{subsec:assessing}).

\begin{table}
\begin{center}
\begin{small}
\begin{tabular}{|l|cc|}
\hline
 & In Sample & Out of Sample \\
\hline
MSFT & $0.781$ & $0.762$ \\ 
INTC & $0.791$ & $0.798$ \\ 
MU & $0.747$ & $0.752$ \\ 
CSCO & $0.802$ & $0.805$ \\ 
ORCL & $0.770$ & $0.770$ \\ 
\hline
GOOG & $0.592$ & $0.581$ \\ 
AMZN & $0.635$ & $0.642$ \\ 
TSLA & $0.602$ & $0.602$ \\ 
PCLN & $0.592$ & $0.583$ \\ 
NFLX & $0.616$ & $0.627$ \\ 
\hline
\end{tabular}
\caption{Area under the ROC curves (see Figure \ref{fig:randLogitROC}) for the logistic regression fits of $\hat{y}$ versus $I$ shown in Figure \ref{fig:randLogit}. The top panel shows the results for large-tick stocks and the bottom panel shows the results for small-tick stocks. For the null model (i.e., $\hat{y}(I)=0.5$ for all $I$), the expected area under the ROC curve is $0.5$.}
\label{tab:randLogitAUC}
\end{small}
\end{center}
\end{table}

For large-tick stocks, the area under the ROC curve varies from about $0.7$ to about $0.8$. For small-tick stocks, the results are weaker, and vary from about $0.6$ to about $0.65$. In both cases, however, these results indicate that $I$ provides a substantial improvement in the out-of-sample predictive power of the binary classifier. To verify that these results are not influenced by over-fitting, we also calculate the area under the corresponding ROC curves for the in-sample fits. In each case, the in-sample values of the statistics in Table \ref{tab:randLogitAUC} are very similar to the corresponding out-of-sample values, which indicates that the predictive power of the logistic regressions is similar for both the training and testing data.

We next address the out-of-sample predictive power of our probabilistic classifier. To do so, we compute the mean squared residual $r_i$ (see Equation (\ref{eq:ri})) between the observed value of $y_i$ and the predicted probability $\hat{y}_i$, across all observations in the testing set (see Table \ref{tab:randLogitMSR}).

\begin{table}
\begin{center}
\begin{small}
\begin{tabular}{|l|cc|cc|}
\hline
 & In Sample & Out of Sample \\
\hline
MSFT & $0.191$ & $0.198$ \\ 
INTC & $0.186$ & $0.183$ \\ 
MU & $0.204$ & $0.202$ \\ 
CSCO & $0.181$ & $0.180$ \\ 
ORCL & $0.195$ & $0.195$ \\ 
\hline
GOOG & $0.244$ & $0.246$ \\ 
AMZN & $0.237$ & $0.235$ \\ 
TSLA & $0.243$ & $0.243$ \\ 
PCLN & $0.244$ & $0.245$ \\ 
NFLX & $0.240$ & $0.239$ \\ 
\hline
\end{tabular}
\caption{Mean squared residual $r_i$ of the logistic regression fits of $\hat{y}$ versus $I$  shown in Figure \ref{fig:randLogit}. The top panel shows the results for large-tick stocks and the bottom panel shows the results for small-tick stocks. For the null model (i.e., $\hat{y}(I)=0.5$ for all $I$), the mean squared residual is $0.25$.}
\label{tab:randLogitMSR}
\end{small}
\end{center}
\end{table}

For large-tick stocks, the mean squared residuals vary from about $0.18$ to about $0.2$. For small-tick stocks, the mean squared residuals vary from about $0.235$ to about $0.245$. For the null model, the mean squared residual of the null model is exactly $1/4$ (see Section~\ref{subsec:assessing}). Therefore, when compared to the null model, the logistic regression fits provide a reduction in mean squared residual of about $20\%$ to $30\%$ for large-tick stocks, and about $2\%$ to $6\%$ for small-tick stocks. For all stocks, the in-sample values of the mean squared residuals are very similar to the corresponding out-of-sample values, which confirms that the logistic regressions do not suffer from over-fitting.

The results in Tables \ref{tab:randLogitAUC} and \ref{tab:randLogitMSR} together present an interesting picture of the out-of-sample performance of our logistic regression fits. In terms of binary classification of the direction of price movements (which we measure by calculating the area under the ROC curve), the logistic regression fits perform well for large-tick stocks and reasonably well for small-tick stocks. In terms of predicting the probability of an upwards mid-price movement (which we measure by calculating the mean squared residuals), the logistic regression fits again perform well for large-tick stocks. For small-tick stocks, however, the results are much weaker, and the logistic regression fits only slightly outperform the null model. Therefore, despite the queue imbalance being a statistically significant predictor of subsequent price movements for all stocks, the improvement in out-of-sample forecasting power that it provides varies considerably across stocks.

\subsection{Local Logistic Regressions}\label{subsec:resultslocallogistic}

In Sections~\ref{subsec:resultslogistic} and \ref{subsec:resultsoos}, we calculated logistic regression fits to estimate a parametric relationship between the queue imbalance and the direction of the subsequent mid-price movement. Although these results are useful for perform fast and simple calculations regarding the statistical significance of the possible relationship between $I$ and $y$, the parametric nature of this approach could obscure the detailed market dynamics that underpin this relationship, because the shape of the fitted logistic regression curves is constrained by the parametric form of the logistic sigmoid function in Equation (\ref{eq:logistic}).

To help address this problem, we now complement our results in Sections~\ref{subsec:resultslogistic} and \ref{subsec:resultsoos} with a semi-parametric approach, by fitting local logistic regression curves to the same data (see Section \ref{subsec:logistic}). In contrast to the logistic regression fits, these semi-parametric fits enable us to consider more carefully the subtle relationship between queue imbalance and mid-price movements, and thereby help to illuminate the market dynamics that underpin our results.

\begin{figure}[!htbp]
\centering
\includegraphics[width=0.9\textwidth]{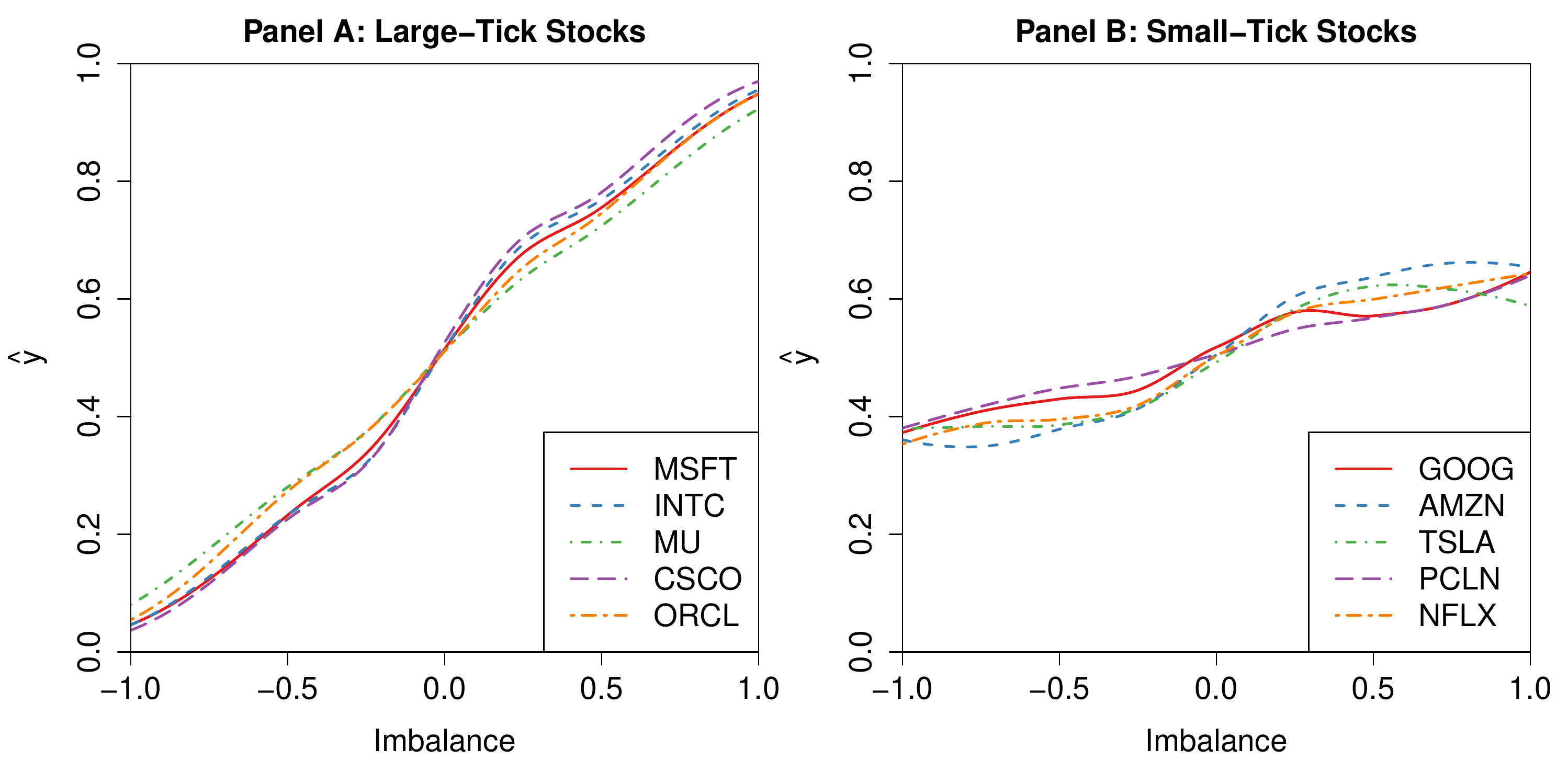}
\caption{Local logistic regression fits of $\hat{y}$ versus $I$. For each curve, we use a tricube weight function and a nearest-neighbour bandwidth of $0.65$. The left panel shows the results for large-tick stocks and the right panel shows the results for small-tick stocks. For full details of our local logistic regression methodology, see Section~\ref{subsec:logistic} and the discussion in the main text.}
\label{fig:randLocalLogit}
\end{figure}

Figure \ref{fig:randLocalLogit} shows our fitted local logistic regression curves for each of the stocks in our sample. For each fit, we use a tricube weight function with a nearest-neighbour bandwidth parameter. To choose this bandwidth parameter, we perform a $5$-fold cross validation within our training set, using the mean squared residual $r_i$ (which we seek to minimize) as our objective function. Although the globally optimal choice of bandwidth parameter varies somewhat across the different stocks in our sample, in all cases it resides between about $0.5$ and about $0.8$. For the results that we present in this section, we use the bandwidth parameter $0.65$. We also repeated all of our calculations for a range of different bandwidth choices between $0.6$ and about $0.7$, and we found that our results were qualitatively similar for all choices in this range.

For large-tick stocks, the local logistic regression curves are approximately monotone increasing functions of $I$, which suggests that larger values of $I_i$ correspond to larger values of $y_i$. Similarly to the logistic regression curves (see Figure \ref{fig:randLogit}), the local logistic regressions predict that the probability of an upward price movement is about $0.8$ to $0.9$ when $I$ is close to $1$. In contrast to the logistic regression curves, however, the local logistic regression curves suggest that the behaviour of the system exhibits 2 different regimes. For values of $I$ between about $-0.25$ and about $0.25$, the $\hat{y}$ curve is quite steep, which indicates that when the bid and ask queues have similar lengths, a relatively small difference in the queue imbalance corresponds to a considerable change in the probability that the next price movement will be upwards. Outside of this region, the steepness of the $\hat{y}$ curve decreases considerably. Therefore, for values of $I$ less than about $-0.25$ or greater than about $0.25$, a further difference in queue imbalance corresponds to a smaller change in the probability that the next price movement will be upwards.

For small-tick stocks, the local logistic regression curves predict that the probability of an upward price movement is about $0.6$ when $I$ is close to $1$. For all small-tick stocks except PCLN, the local logistic regression curves are non-monotonic in $I$. This result is rather puzzling, because it suggests that there are cases when a weaker imbalance increases the probability of an upward mid-price movement. This counter-intuitive finding brings into question whether the fitted local logistic regressions $\hat{y}$ really detect a meaningful relationship, or simply over-fit to noise.

\begin{figure}[!htbp]
\centering
\includegraphics[width=0.9\textwidth]{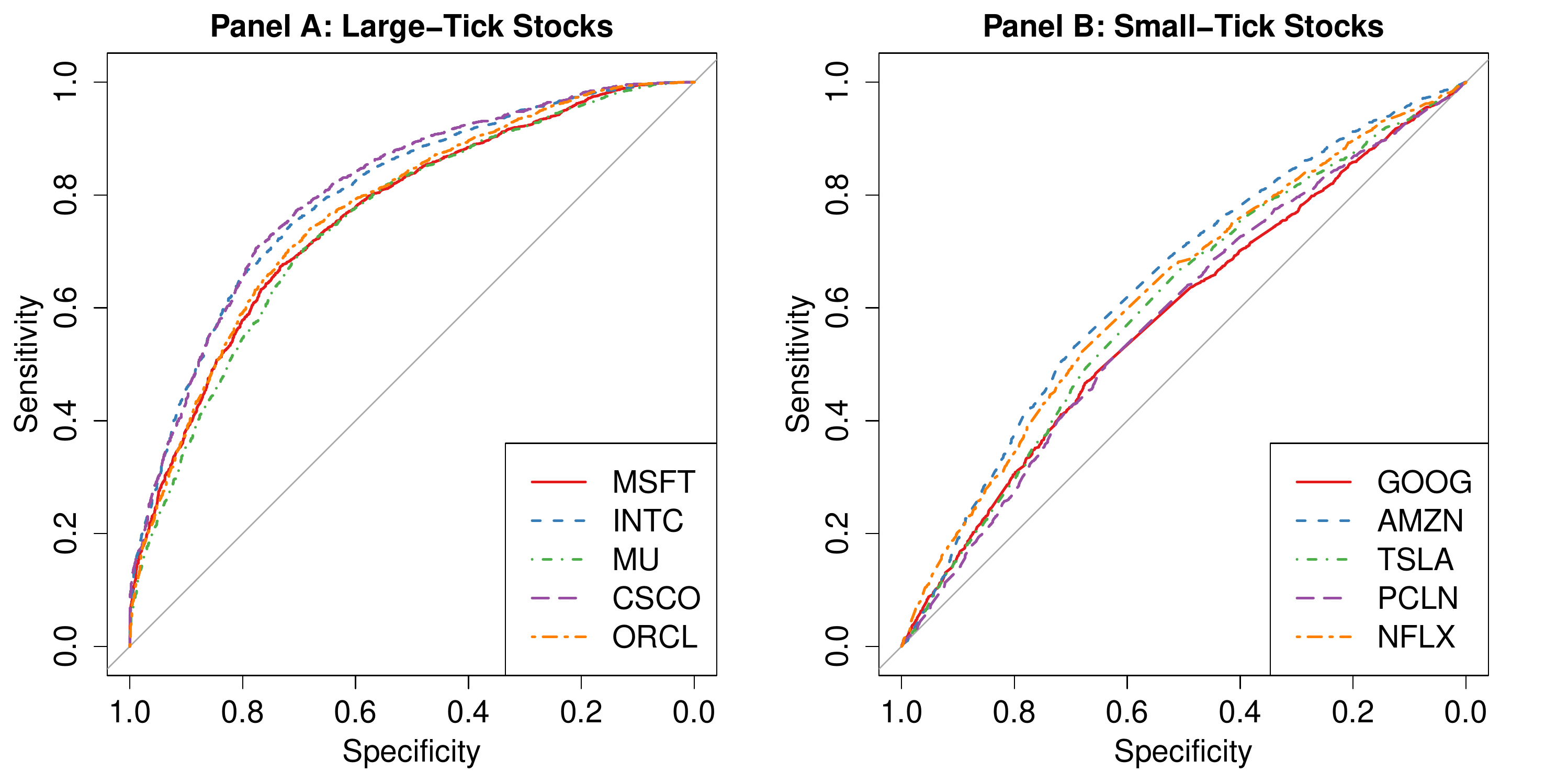}
\caption{Receiver operating characteristic (ROC) curves for the out-of-sample predictive power of the local logistic regression fits of $\hat{y}$ versus $I$. The left panel shows the results for large-tick stocks and the right panel shows the results for small-tick stocks. The grey line in each plot denotes the expected performance of the null model, which assumes that the probability of an upward price movement is always equal to $1/2$, irrespective of the queue imbalance.}
\label{fig:randLocalLogitROC}
\end{figure}

To address this question, we again consider the out-of-sample predicted power of the local logistic regression curves for performing binary classification and probabilistic classification. Figure~\ref{fig:randLocalLogitROC} shows the out-of-sample ROC curves (see Section~\ref{subsec:logistic}) for each of our local logistic regression fits, together with the corresponding ROC curve for the null model of $\hat{y}(I)=1/2$ for all $I$.

For each stock, the out-of-sample ROC curve for the local logistic regressions is very similar to the corresponding ROC curve for the logistic regression. In all cases, the ROC curve lies above the grey line, which indicates that the local logistic regression fits outperform the out-of-sample predictive power of the null model at all levels of specificity.

To quantify the strength of this increase in predictive power, and to consider the out-of-sample predictive power of the local logistic regression curves for performing probabilistic classification, we again calculate the area under the ROC curve (see the first two columns of Table~\ref{tab:randLocalLogitAUC}) and the mean squared residual $r_i$ (see the final two columns of Table~\ref{tab:randLocalLogitAUC}).

\begin{table}
\begin{center}
\begin{small}
\begin{tabular}{|l|cc|cc|}
\hline
 & \multicolumn{2}{|c|}{Area Under ROC Curve} & \multicolumn{2}{|c|}{Mean Squared Residual} \\
\cline{2-5}
 & In Sample & Out of Sample & In Sample & Out of Sample \\
\hline
MSFT & $0.781$ & $0.762$ & $0.190$ & $0.198$ \\ 
INTC & $0.791$ & $0.798$ & $0.185$ & $0.183$ \\ 
MU & $0.747$ & $0.752$ & $0.204$ & $0.202$ \\ 
CSCO & $0.802$ & $0.805$ & $0.181$ & $0.179$ \\ 
ORCL & $0.770$ & $0.770$ & $0.195$ & $0.195$ \\ 
\hline
GOOG & $0.592$ & $0.581$ & $0.244$ & $0.246$ \\ 
AMZN & $0.636$ & $0.642$ & $0.236$ & $0.235$ \\ 
TSLA & $0.602$ & $0.603$ & $0.242$ & $0.242$ \\ 
PCLN & $0.592$ & $0.583$ & $0.244$ & $0.245$ \\ 
NFLX & $0.616$ & $0.627$ & $0.240$ & $0.239$ \\ 
\hline
\end{tabular}
\caption{Statistics describing the predictive power of the local logistic regression fits of $\hat{y}$ versus $I$. The first two columns show the area under the ROC curve and the second two columns show the mean squared residual of the fits. The top panel shows the results for large-tick stocks and the bottom panel shows the results for small-tick stocks. For the null model (i.e., $\hat{y}(I)=0.5$ for all $I$), the expected area under the ROC curve is $0.5$ and the mean squared residual is $0.25$. For a full discussion of our local logistic regression methodology, see Section~\ref{subsec:logistic} and the discussion in the main text.}
\label{tab:randLocalLogitAUC}
\end{small}
\end{center}
\end{table}

Similarly to our results in Tables~\ref{tab:randLogitAUC} and \ref{tab:randLogitMSR}, the in-sample values of the statistics in Table \ref{tab:randLocalLogitAUC} are very similar to the corresponding out-of-sample values, which indicates that the predictive power of the local logistic regressions is similar for both the training and testing data. Therefore, the unusual shape of the $\hat{y}$ functions in Figure~\ref{fig:randLocalLogitROC} is not a consequence of over-fitting to noise.

By comparing the performance measures for the logistic regressions in Tables~\ref{tab:randLogitAUC} and \ref{tab:randLogitMSR} to the corresponding performance measures for the local logistic regressions (see Table~\ref{tab:randLocalLogitAUC}), we are able to quantify the differences between these different approaches. Interestingly, the values in Tables~\ref{tab:randLogitAUC} and \ref{tab:randLogitMSR} are all very similar to the corresponding values in Table~\ref{tab:randLocalLogitAUC}. This suggests that the performance of these two different approaches is quite similar. Detailed comparisons of these tables reveal that the local logistic regressions slightly outperform the logistic regressions in some cases, but these improvements are all quite small. We provide a more detailed comparison of our results from logistic regression and local logistic regression in Section~\ref{sec:discussion}.

\section{Discussion}\label{sec:discussion}

Our results in Section \ref{sec:results} illustrate the existence of a statistically significant relationship between the queue imbalance and the direction of the subsequent mid-price movement. For large-tick stocks, the relationship depends quite strongly on the queue imbalance (see upper rows of Table~\ref{tab:randLogit}) and provides a considerable improvement in out-of-sample predictive power in terms of both binary classification (see Table~\ref{tab:randLogitAUC}) and probabilistic classification (see Table~\ref{tab:randLogitMSR}). For small-tick stocks, the relationship depends less strongly on the queue imbalance (see lower rows of Table~\ref{tab:randLogit}), and the improvement in out-of-sample predictive power is more moderate (see Table~\ref{tab:randLocalLogitAUC}).

Among the stocks in our sample, the weakest out-of-sample performance that we observe occurs in the probabilistic classification for GOOG, for which our fits outperform the null model by about $2\%$. Although this number is certainly small compared to the performance that we achieve for other stocks, it is important to remember that many practitioners invest huge sums of money to improve their trading strategies by tiny fractions of a percentage point. Therefore, even this very moderate performance for GOOG could be economically significant for some market participants. Moreover, both of our performance measures are unconditional averages that make forecasts for all data points, many of which correspond to situations where the imbalance is small (see Figure~\ref{fig:randHist}), and therefore where the predictability of the mid-price movement is weak. If we considered only the situations in which the queue imbalance was close to $\pm 1$, then the out-of-sample performance of our estimators would improve considerably. This observation is particularly important from a practical standpoint, because some practitioners may only be interested in forecasting when the ability to do so is likely to be strong, and may therefore simply abstain from trading when $I \approx 0$.

Similarly, it seems reasonable to assume that $I$ is less informative when both $n^b(b(t),t)$ and $n^a(a(t),t)$ are small, because the arrival of a single buy (respectively, sell) market order is likely to cause the mid price to increase (respectively, decrease). Therefore, if we considered only the situations in which both $n^b(b(t),t)$ and $n^a(a(t),t)$ are considerably larger than 0, then the out-of-sample performance of our estimators would again improve considerably.

It is interesting to consider why our results for large-tick stocks are so different from our results for small-tick stocks. We believe that the answer to this puzzle lies in the underlying market microstructure. Recall from Section \ref{sec:lobs} that in an LOB, there are usually two ways for $m(t)$ to change: by a new limit order arriving inside the bid--ask spread, or by one of the bid or ask queue lengths depleting to 0. However, the mean bid--ask spread for the large-tick stocks in our sample is very close to its minimum possible value of the platform's tick size, $s(t)=\pi=\$0.01$ (see Table~\ref{tab:summaries}). This behaviour has an important consequence for LOB dynamics, because it removes the possibility that a new limit order will arrive inside the bid--ask spread, and thereby eliminates one of the two possible reasons for changes in $m(t)$. It is therefore reasonable to believe that the queue imbalance (which quantifies the relative lengths of the bid and ask queues) will provide stronger predictive power when $s(t)=\pi$, because the probability of an upwards price movement is governed only by the probability that the ask queue depletes before the bid queue (which, in turn, depends directly on the queue lengths), and not on the probability that a new buy limit order arrives inside the spread.

Even when $s(t)>\pi$, there are still strong reasons for why market participants may behave differently for small-tick and large-tick stocks. Similarly to most other LOBs, the Nasdaq platform operates a price--time priority rule (see Section~\ref{sec:data}), by which priority is given to the active buy (respectively, sell) orders with the highest (respectively, lowest) price, and ties are broken by selecting the active order with the earliest submission time. When $s(t)>\pi$, any market participant has the opportunity to submit a buy (respectively, sell) limit order with higher priority than any others in the LOB, simply by choosing the price of this order to be one tick higher than $b(t)$ (respectively, lower than $a(t)$). Therefore, the tick size $\pi$ determines the cost of ``buying'' priority in the LOB. For large-tick stocks, this cost is relatively high, so many traders choose to submit new limit orders that wait in the queues at the best quotes, and the typical sizes of $n^b(b(t),t)$ and $n^a(a(t),t)$ are large. For small-tick stocks, this cost is relatively low, so many traders choose to submit new limit orders inside the bid--ask spread, and the typical sizes of $n^b(b(t),t)$ and $n^a(a(t),t)$ are small. As noted above, the predictive power of $I$ is likely to be larger in situations where $n^b(b(t),t)$ and $n^a(a(t),t)$ are larger. In this way, the stronger out-of-sample performance for large-tick stocks can similarly be attributed to the longer queue lengths that typically occur for these stocks.

For some stocks in our sample, we find a weak but statistically significant asymmetry in the fitted logistic regressions, which manifests as a non-zero value of the intercept $x_0$ (see Table~\ref{tab:randLogit}). For these stocks, this result suggests that even when the imbalance is slightly less than 0, the probability that the next mid-price movement will be upwards is greater than $1/2$. We propose two possible explanations for this finding. First, the arrival of exogenous news may cause traders to change their trading behaviours, irrespective of the queue imbalance. For example, if a trader receives news that a given company's earnings have outperformed expectations, then he/she may submit a large market order to buy the stock (and thereby cause an increase in mid price), even if the current queue imbalance is negative. We note that all stocks for which we find $x_0$ to be statistically significantly positive underwent considerable price increases during 2014, which is consistent with this hypothesis of an exogenous buying pressure. Second, some strategic liquidity providers may implement complex strategies that skew the queue imbalance via an asymmetric submission of limit orders. For example, if a strategic liquidity provider fears the possibility of a strong downward price movement, then he/she may choose to submit fewer sell limit orders than buy limit orders, even in the absence of any information about the likely future value of the asset. In a recent empirical study of strategic liquidity provision on Nasdaq, \citet{Bonart:2015strategic} found strong evidence to suggest that liquidity providers implement strategies that created imbalanced net order flow at the best quotes. Although it is difficult to test these theories directly, we believe that both of these explanations are likely to contribute to the behaviour that we observe.

In addition to our logistic regression fits, we also perform local logistic regression fits on the data. By comparing the entries in Tables~\ref{tab:randLogitAUC} and \ref{tab:randLogitMSR} to the corresponding entries in Table \ref{tab:randLocalLogitAUC}, it is possible to compare the performance of the logistic regression and local logistic regression fits. Most of the entries in Tables~\ref{tab:randLogitAUC} and \ref{tab:randLogitMSR} are equal to the corresponding entries in Table \ref{tab:randLocalLogitAUC}, even up to the third decimal place. This implies that the performance of the two methods is very similar. In a small number of cases, however, the local logistic regression fits slightly outperform the corresponding logistic regression fits.

Both approaches have benefits and drawbacks. Performing logistic regression is much less computationally intensive than performing local logistic regression, and the full fit of the logistic regression model consists of just 2 scalar values, $x_0$ and $x_1$. Saving the fitted logistic regression curve to a computer hard disk  therefore requires very little storage space. By contrast, saving the fitted local logistic regression curve requires saving a full copy of the training data, which can be very large. However, local logistic regression has the important benefit of providing more detailed information about the underlying LOB dynamics, because the fitted regression curve is not constrained by the parametric form of the logistic sigmoid function. Therefore, careful analysis of the local logistic regression fits can provide deeper understanding of the results.


\section{Conclusions and Outlook}\label{sec:conclusions}

In this paper, we have presented an empirical study of whether the queue imbalance $I$ provides significant predictive power for the direction $y$ of the next mid-price movement. We used data describing the LOB activity for each of 10 liquid stocks on Nasdaq during 2014 to fit logistic regression curves that enabled us to perform both binary classification and probabilistic classification of $y$, given $I$. For all 10 stocks in our sample, we found that our logistic regressions identified a strongly statistically significant relationship between $I$ and $y$.

For the large-tick stocks in our sample, our logistic regression fits provide a considerable improvement in both binary and probabilistic classification. For the small-tick stocks, we found that the increase in predictive power was more moderate, particularly for probabilistic classification. We argued that the reason for these differences was the differences in underlying market microstructure. We also performed local logistic regression fits on the same data, and found that this semi-parametric approach slightly outperforms the logistic regression fits, at the expense of being more computationally expensive.

In addition to these practical benefits, our results also highlight many possible avenues for future research. Throughout this paper, we have chosen to measure the queue imbalance via the quantity $I$, according to Equation (\ref{eq:I}), sampled at a time chosen uniformly at random between subsequent changes of $m(t)$. Although we identify a statistically significant relationship between $I$ and $y$, there are many other possible ways to measure queue imbalance in an LOB. For example, we could measure the value of $I$ immediately after or immediately before each mid-price change, to examine how its predictive power varies according to the length of time that elapses before the next price change. Similarly, it would be interesting to see whether an alternative definition of the queue imbalance could provide stronger out-of-sample predictive power than the quantity $I$ that we used for this study. Other possibilities could include quantities such as $\log(n^b(b(t),t)/n^a(a(t),t))$, or even simply $n^b(b(t),t)-n^a(a(t),t)$. Moreover, we have only studied the imbalance between the best bid and ask queues. It is possible that the predictive power of our approach could be improved by also incorporating other statistics about the lengths or imbalance of other queues deeper into the LOB.

Our results also raise interesting questions about the predictability of price movements on longer time scales. For example, does $I$ provide useful information about the direction of price movements further into the future? If so, how does this predictive power diminish over time? And how do price movements remain unpredictable on longer timescales, given that we show them to be quite predictable in our one-step-ahead framework?

Understanding the relationship between queue imbalance and subsequent price movements is also an important theoretical question. It would therefore be interesting to build models for how imbalance could evolve over time, and could thereby affect the subsequent evolution of LOB state. In comparison to modelling the full state of an LOB (which is an extremely high-dimensional problem), the simple, one-dimensional nature of $I(t)$ makes modelling its temporal evolution an attractive task.  We aim to address these and many other questions about the predictive power of queue imbalance in our future work.

\section*{Acknowledgements}We thank Jean-Philippe Bouchaud, Rama Cont, Jonathan Donier, Till Hoffmann, Nick Jones, Julien Kockelkoren, Charles-Albert Lehalle, and Douglas Machado for useful discussions. We thank Jonas Haase and Ruihong Huang for technical support. Martin D. Gould gratefully acknowledges support from the James S. McDonnell Foundation and Julius Bonart gratefully acknowledges support from CFM.

\bibliographystyle{plainnat}
\bibliography{../dphilLOBbib}

\begin{thebibliography}{26}
\providecommand{\natexlab}[1]{#1}
\providecommand{\url}[1]{\texttt{#1}}
\expandafter\ifx\csname urlstyle\endcsname\relax
  \providecommand{\doi}[1]{doi: #1}\else
  \providecommand{\doi}{doi: \begingroup \urlstyle{rm}\Url}\fi

\bibitem[Avellaneda et~al.(2011)Avellaneda, Reed, and
  Stoikov]{Avellaneda:2011forecasting}
M.~Avellaneda, J.~Reed, and S.~Stoikov.
\newblock Forecasting prices from level-{I} quotes in the presence of hidden
  liquidity.
\newblock \emph{Algorithmic Finance}, 1\penalty0 (1):\penalty0 35--43, 2011.

\bibitem[Bonart and Gould(2015)]{Bonart:2015strategic}
J.~Bonart and M.~D. Gould.
\newblock Strategic liquidity provision in a limit order book.
\newblock \emph{arXiv:1511.04116}, 2015.

\bibitem[Bouchaud et~al.(2009)Bouchaud, Farmer, and Lillo]{Bouchaud:2009digest}
J.~P. Bouchaud, J.~D. Farmer, and F.~Lillo.
\newblock How markets slowly digest changes in supply and demand.
\newblock In T.~Hens and K.~R. Schenk-Hopp{\'{e}}, editors, \emph{Handbook of
  Financial Markets: Dynamics and Evolution}, pages 57--160. North--Holland,
  Amsterdam, The Netherlands, 2009.

\bibitem[Bradley(1997)]{Bradley:1997use}
A.~P. Bradley.
\newblock The use of the area under the {ROC} curve in the evaluation of
  machine learning algorithms.
\newblock \emph{Pattern Recognition}, 30\penalty0 (7):\penalty0 1145--1159,
  1997.

\bibitem[Cao et~al.(2009)Cao, Hansch, and Wang]{Cao:2009information}
C.~Cao, O.~Hansch, and X.~Wang.
\newblock The information content of an open limit-order book.
\newblock \emph{Journal of Futures Markets}, 29\penalty0 (1):\penalty0 16,
  2009.

\bibitem[Cartea et~al.(2015)Cartea, Donnelly, and
  Jaimungal]{Cartea:2015enhancing}
{\'A}.~Cartea, R.~F. Donnelly, and S.~Jaimungal.
\newblock Enhancing trading strategies with order book signals.
\newblock \emph{Working Paper{,} SSRN eLibrary ID 2668277}, 2015.

\bibitem[Chakraborti et~al.(2011)Chakraborti, Toke, Patriarca, and
  Abergel]{Chakraborti:2011empirical}
A.~Chakraborti, I.~M. Toke, M.~Patriarca, and F.~Abergel.
\newblock Econophysics review {I}: {E}mpirical facts.
\newblock \emph{Quantitative Finance}, 11\penalty0 (7):\penalty0 991--1012,
  2011.

\bibitem[Cont(2001)]{Cont:2001empirical}
R.~Cont.
\newblock Empirical properties of asset returns: stylized facts and statistical
  issues.
\newblock \emph{Quantitative Finance}, 1\penalty0 (2):\penalty0 223--236, 2001.

\bibitem[Cont and De~Larrard(2013)]{Cont:2013price}
R.~Cont and A.~De~Larrard.
\newblock Price dynamics in a {M}arkovian limit order market.
\newblock \emph{SIAM Journal on Financial Mathematics}, 4\penalty0
  (1):\penalty0 1--25, 2013.

\bibitem[Cont et~al.(2010)Cont, Stoikov, and Talreja]{Cont:2010stochastic}
R.~Cont, S.~Stoikov, and R.~Talreja.
\newblock A stochastic model for order book dynamics.
\newblock \emph{Operations Research}, 58\penalty0 (3):\penalty0 549--563, 2010.

\bibitem[Farmer et~al.(2005)Farmer, Patelli, and Zovko]{Farmer:2005predictive}
J.~D. Farmer, P.~Patelli, and I.~I. Zovko.
\newblock The predictive power of zero intelligence in financial markets.
\newblock \emph{Proceedings of the National Academy of Sciences of the United
  States of America}, 102\penalty0 (6):\penalty0 2254--2259, 2005.

\bibitem[Farmer et~al.(2006)Farmer, Gerig, Lillo, and Mike]{Farmer:2006market}
J.~D. Farmer, A.~N. Gerig, F.~Lillo, and S.~Mike.
\newblock Market efficiency and the long-memory of supply and demand: {I}s
  price impact variable and permanent or fixed and temporary?
\newblock \emph{Quantitative Finance}, 6\penalty0 (2):\penalty0 107--112, 2006.

\bibitem[Gar\`{e}che et~al.(2013)Gar\`{e}che, Disdier, Kockelkoren, and
  Bouchaud]{Gareche:2013fokker}
A.~Gar\`{e}che, G.~Disdier, J.~Kockelkoren, and J.~P. Bouchaud.
\newblock Fokker--{P}lanck description for the queue dynamics of large tick
  stocks.
\newblock \emph{Physical Review E}, 88\penalty0 (3):\penalty0 032809, 2013.

\bibitem[Gould et~al.(2013)Gould, Porter, Williams, McDonald, Fenn, and
  Howison]{Gould:2013limit}
M.~D. Gould, M.~A. Porter, S.~Williams, M.~McDonald, D.~J. Fenn, and S.~D.
  Howison.
\newblock Limit order books.
\newblock \emph{Quantitative Finance}, 13\penalty0 (11):\penalty0 1709--1742,
  2013.

\bibitem[Hanley and McNeil(1982)]{Hanley:1982meaning}
J.~A. Hanley and B.~J. McNeil.
\newblock The meaning and use of the area under a receiver operating
  characteristic curve.
\newblock \emph{Radiology}, 143\penalty0 (1):\penalty0 29--36, 1982.

\bibitem[Hastie et~al.(2010)Hastie, Tibshirani, and
  Friedman]{Hastie:2009elements}
T.~Hastie, R.~Tibshirani, and J.~Friedman.
\newblock \emph{The Elements of Statistical Learning}.
\newblock Springer, New York, NY, USA, 2010.

\bibitem[Hosmer and Lemeshow(2004)]{Hosmer:2004applied}
D.~W. Hosmer and S.~Lemeshow.
\newblock \emph{Applied Logistic Regression}.
\newblock Wiley, New York, NY, USA, 2004.

\bibitem[Huang et~al.(2015)Huang, Lehalle, and Rosenbaum]{Huang:2015simulating}
W.~Huang, C.~A. Lehalle, and M.~Rosenbaum.
\newblock Simulating and analyzing order book data: {T}he queue-reactive model.
\newblock \emph{Journal of the American Statistical Association}, 110\penalty0
  (509):\penalty0 107--122, 2015.

\bibitem[{Knight Capital Group}(2015)]{HotspotGUIUserGuide}
{Knight Capital Group}.
\newblock Retrieved 14 April 2015 from
  \url{http://www.hotspotfx.com/download/userguide/HSFX/HSFX_UserGuide_wrapper.html},
  2015.

\bibitem[Loader(2006)]{Loader:2006local}
C.~Loader.
\newblock \emph{Local Regression and Likelihood}.
\newblock Springer, New York, NY, USA, 2006.

\bibitem[McCullagh and Nelder(1989)]{Mccullagh:1989generalized}
P.~McCullagh and J.~A. Nelder.
\newblock \emph{Generalized Linear Models}, volume~37.
\newblock Chapman and Hall, London, UK, 1989.

\bibitem[Mike and Farmer(2008)]{Mike:2008empirical}
S.~Mike and J.~D. Farmer.
\newblock An empirical behavioral model of liquidity and volatility.
\newblock \emph{Journal of Economic Dynamics and Control}, 32\penalty0
  (1):\penalty0 200--234, 2008.
\newblock ISSN 0165-1889.

\bibitem[Ro\c{s}u(2009)]{Rosu:2009dynamic}
I.~Ro\c{s}u.
\newblock A dynamic model of the limit order book.
\newblock \emph{Review of Financial Studies}, 22\penalty0 (11):\penalty0
  4601--4641, 2009.

\bibitem[Smith et~al.(2003)Smith, Farmer, Gillemot, and
  Krishnamurthy]{Smith:2003statistical}
E.~Smith, J.~D. Farmer, L.~Gillemot, and S.~Krishnamurthy.
\newblock Statistical theory of the continuous double auction.
\newblock \emph{Quantitative Finance}, 3\penalty0 (6):\penalty0 481--514, 2003.

\bibitem[Stoikov and Waeber(2015)]{Stoikov:2015reducing}
S.~Stoikov and R.~Waeber.
\newblock Reducing transaction costs with low-latency trading algorithms.
\newblock \emph{Working Paper{,} SSRN eLibrary ID 2661618}, 2015.

\bibitem[Yang and Zhu(2015)]{Yang:2015reduced}
T.~W. Yang and L.~Zhu.
\newblock A reduced-form model for level-1 limit order books.
\newblock \emph{arXiv:1508.07891}, 2015.

\end{thebibliography}

\end{document}